\documentclass[openany]{aa}  

\usepackage{hyperref}  
\hypersetup{colorlinks=true,linkcolor=[rgb]{1.,0.2,0.2},citecolor=[rgb]{0.1,0.4,1.},filecolor=[rgb]{0.7,0.2,0.2},urlcolor=[rgb]{0.7,0.2,0.2}}

\usepackage{graphicx}
\usepackage{txfonts}
\usepackage{lipsum}
\usepackage{subcaption}         % necessary for continued figures, example in section 3
\usepackage{lscape}             % to rotate a single page table, example in appendix.
\usepackage{placeins}           % useful with \FloatBarrier, to keep 
\newcommand{\msun}{M$_{\sun}$}

\newcommand{\kms}{km~s$^{-1}$}
\newcommand{\ha}{H$\alpha$}
\newcommand{\hb}{H$\beta$}
\newcommand{\micron}{$\mu$m}
\begin{document}

\title{Spatially resolved metallicity and ionization in the merging system Gz9p3 at z=9.3}

\author{Arjan Bik\inst{1}\corrauth{arjan.bik@astro.su.se}
\and Javier Álvarez-M\'arquez\inst{2}
\and Alejandro Crespo Gómez \inst{2,3}
\and Luis Colina \inst{2}
\and Pablo G. Pérez-González \inst{2}
\and Göran Östlin \inst{1}
\and Carmen Blanco Prieto \inst{2}
\and Jens Melinder \inst{1}
\and Danial  Langeroodi \inst{4}
\and Gillian Wright \inst{5}
\and Hiddo S. B. Algera \inst{6}
\and Leindert A. Boogaard \inst{7}
\and Karina Caputi \inst{8}
\and Steven Gillman \inst{9,10}
\and Thomas Greve \inst{10,9}
\and Jens Hjorth \inst{4}
\and Edoardo Iani \inst{11}
\and Sarah Kendrew \inst{12}
\and Alvaro Labiano Ortega \inst{2,13}
\and Michele Perna \inst{2}
\and Carlota Prieto Jimenez \inst{2}
\and John Pye \inst{14}
\and Pierluigi Rinaldi \inst{3}
\and  Paul van der Werf \inst{7}
\and Fabian Walter \inst{15}
\and Florian Pei{\ss}ker \inst{16}
\and Andreas Eckart \inst{16}
\and Thomas Henning \inst{15}
\and Manuel G\"udel \inst{17,18,19}
}

\institute{Department of Astronomy, Stockholm University, Oscar Klein Centre, AlbaNova University Centre, 106 91 Stockholm, Sweden
\and Centro de Astrobiología (CAB), CSIC-INTA, Ctra. de Ajalvir km 4, Torrejón de Ardoz, E-28850, Madrid, Spain
\and Space Telescope Science Institute (STScI), 3700 San Martin Drive, Baltimore, MD, 21218, USA
\and DARK, Niels Bohr Institute, University of Copenhagen, Jagtvej 155A, 2200 Copenhagen, Denmark
\and UK Astronomy Technology Centre, Royal Observatory Edinburgh, Blackford Hill, Edinburgh EH9 3HJ, UK
\and Institute of Astronomy and Astrophysics, Academia Sinica, 11F of Astronomy-Mathematics Building, No.1, Sec. 4, Roosevelt Rd, Taipei 106319, Taiwan, R.O.C.
\and Leiden Observatory, Leiden University, PO Box 9513, NL-2300 RA Leiden, The Netherlands
\and Kapteyn Astronomical Institute, University of Groningen, P.O. Box 800, 9700 AV Groningen, The Netherlands
\and DTU Space, Technical University of Denmark, Elektrovej 327, 2800 Kgs. Lyngby, Denmark
\and Cosmic Dawn Center (DAWN), Denmark
\and Institute of Science and Technology Austria, Am Campus 1, 3400 Klosterneuburg, Austria
\and European Space Agency (ESA), ESA Office, Space Telescope Science Institute, 3700 San Martin Drive, Baltimore, MD 21218, USA
\and Telespazio UK for the European Space Agency, ESAC, Camino Bajo del Castillo s/n, E-28692 Villanueva de la Cañada, Madrid, Spain
\and School of Physics \& Astronomy, Space Park Leicester, University of Leicester, 92 Corporation Road, Leicester LE4 5SP, UK
\and Max-Planck-Institut für Astronomie, Königstuhl 17, 69117 Heidelberg, Germany
\and I.Physikalisches Institut der Universit\"at zu K\"oln, Z\"ulpicher Str. 77, 50937 K\"oln, Germany
\and Dept. of Astrophysics, University of Vienna, T\"urkenschanzstr. 17, A-1180 Vienna, Austria
\and ETH Z\"urich, Institute for Particle Physics and Astrophysics, Wolfgang-Pauli-Str. 27, 8093 Z\"urich, Switzerland
\and ASTRON, Netherlands Institute for Radio Astronomy, Oude Hoogeveensedijk 4, 7991 PD Dwingeloo, The Netherlands
}

%15/01

% to ask: Michele Perna; done 
% to ask: MIRI CoPIs: 
% Manuel Guedel
% Thomas Henning
% Pierre Olivier Lagage
% Tom Ray
% Ewmine van Dishoeck

% ackn: Almudena Alonso Herrero

\date{Received / Accepted }

  \abstract
   {Studying the interstellar medium (ISM) in merging high‑redshift galaxies is crucial for understanding early galaxy assembly, star formation, and black hole growth, predicted by hierarchical $\Lambda$CDM models. Deep imaging and spatially resolved spectroscopy with JWST enable unprecedented insight into these processes, even for galaxies in the Epoch of Reionization.}
   {We present NIRSpec and MIRI integral field spectroscopy and MIRI imaging of the merging galaxy Gz9p3 at z=9.3 observing the UV and optical rest-frame emission showing a clumpy morphology in the continuum as well as line emission covering the entire galaxy over a range of 5 kpc from the central clump to the tail region.} 
   {We analyze the integrated spectrum as well as different apertures in the galaxy allowing a spatially resolved characterization of the ionized ISM of this merging galaxy. We compare our measurements with archival NIRCam imaging as well as ALMA data.}
   {We measure a total star formation rate of 13.4 $\pm$ 1.8 M$_{\odot}$ yr$^{-1}$, a metallicity of 12+log(O/H)= 7.84 $\pm$ 0.05 and a $\xi_{ion}$ = 25.4 $\pm 0.1$ erg$^{-1}$ Hz and a burstiness parameter of 0.9 $\pm$ 0.1 for the integrated spectrum. We find large spatial differences in these parameters between the central clump and the tail region. While the optical [OIII] emission peaks in the main galaxy, the far-infrared [OIII] emission peaks towards the tidal tail, indicating different physical conditions in the ISM of the tail and main galaxy.}
   {This study presents the spatially resolved ISM analyses of a galaxy at z>9, revealing nebular line emission and strong spatial variations in star formation, metallicity, physical conditions, and ionizing efficiency. The results indicate a recent, metal‑poor starburst in a tail alongside a more evolved, enriched central clump with evidence for extreme excitation. This demonstrates the power of spatially resolved JWST spectroscopy of galaxies in the Epoch of Reionization.}


   \maketitle
      \nolinenumbers
\section{Introduction} \label{sec:intro}

Probing the physical conditions of the interstellar medium (ISM) in galaxies at high redshift is essential for understanding the earliest phases of galaxy assembly and formation. In the first billion years after the Big Bang, galaxies experience rapid growth, intense star formation, and the onset of chemical enrichment \citep[e.g.][]{Adamoreview25}. JWST has provided a revolution in the study of the earliest galaxies, both via large imaging surveys as well spectroscopic programs \citep[and references therein]{Eisenstein26,Stark26}.

The $\Lambda$CDM model predicts that galaxies grow over time \citep{White78} via accretion and merging of smaller galaxies, where the fraction of merging galaxies is higher at high redshift \citep{Mortlock15,Duan25}. High‑redshift mergers are expected to trigger intense starbursts, drive morphological transformations, and efficiently channel gas toward galactic nuclei, thereby linking galaxy growth to the early evolution of supermassive black holes and active galactic nuclei \citep{Barnes92,Hopkins06}. They provide a uniquely powerful probe of the physical processes that shaped the present‑day galaxy population.

With the availability of spatially resolved spectroscopy with NIRSpec-IFU and MIRI-MRS, the sensitivity of JWST even allows spatially resolved studies of galaxies in the Epoch of Reionization (EoR), as young as $\sim$ 400 -- 500 Myrs after the Big Bang by revealing variations in star formation and ISM properties \citep[e.g.][]{Bunker23,Marconcini24,Arribas23,Alvarez23,Alvarez24,Messa25,Zamora25}. In this paper we focus on a detailed analysis of the ISM of the galaxy  Gz9p3, a merging galaxy observed only 500 Myrs after the Big Bang.

Gz9p3 is observed at a redshift of z=  9.3 and is located behind the periphery of the galaxy cluster Abell 2744. The galaxy was identified based on the HST Frontier field observations \citep{Castellano16,Yue18} as F105W and F125W dropout, making it one of the z$\sim$9 candidates already identified by HST observations. JWST NIRCam observations of the  GLASS team \citep[under the name DHz1]{Castellano23} as well as the UNCOVER team \citep{Atek23} confirmed its high-redshift nature and derived photometric redshifts between 9.4 and 9.8 \citep{Atek23}. Spectroscopic observations with NIRSpec-MSA using the high resolution grating by \citet{Boyett24} measured the redshift of Gz9p3 to be 9.3127 $\pm$ 0.0002. 

The same authors present NIRCam multi-wavelength imaging of the galaxy and performed spatially resolved SED fitting. The imaging reveals a clumpy galaxy consisting of a central region with two UV bright clumps region of very young stars ($\sim$ 10 Myr) surrounded by a spatially extended old stellar population (120 $\pm$ 20 Myr). Additionally, this galaxy exhibits an elongated clumpy tidal tail extending out to $\sim$ 5 kpc from the central region. Based on this morphology and combined with the spectral analysis \citet{Boyett24} concluded that it is an interacting massive (M$_{\star}$ =  1.6$^{+0.5}_{-0.4} \times 10^{9}$ \msun) galaxy.

Based on the [NeIII] over [OII] (Ne3O2) line ratio a metallicity of 12 + log(O/H) = 7.6 $\pm$ 0.5 was derived. Observations with  NIRSpec-MSA with the PRISM of Gz9p3 revealed a high [OIII] over \hb\ (R3 $\sim 1$) ratio, raising the suggestion that an AGN could be present in this galaxy \citep{Fujimoto24}. Recently deep ALMA observations revealed spatially resolved [OIII] 88\micron\ emission \citep{Algera25}, where the [OIII] 88\micron\  emission peaks the strongest towards the bridge between the central clump and the tail. They do not detect the galaxy in the continuum emission and place an upper limit for the dust mass of 10$^{6}$ M$_{\odot}$ assuming a temperature of 50 K.

In this paper we present deep NIRSpec-IFU, MIRI-MRS and MIRIM observations of Gz9p3 enabling for the first time a spatially resolved analysis of the ISM of this interactive galaxy. The NIRSpec and MIRI spectra cover the full UV and optical restframe allowing us derive the ionization, star formation rate, ionizing photon efficiency, extinction and metallicity in a spatially resolved manner and look for variations across the galaxy. This will enable us to put constraints on the nature of the ionizing sources and the progress of star formation in this galaxy.

This paper is organized as follows, in Sect. \ref{sec:observations} we present the NIRSpec and MIRI spectroscopy and imaging, as well as the auxiliary data used in this study. In Sect. \ref{sec:results} we present the spectra and spatially resolved data and extract the measurements from  the data and in sect. \ref{sec:physical_properties} we derive the physical properties of the galaxies based on the emission line measurements. In Sect. \ref{sec:discussion} we place the results in context with other high-z studies and the paper ends with conclusions (Sect. \ref{sec:conclusions}). 

We adopt a lensing magnification of $\mu = 1.66 \pm 0.02$ as in \citet{Castellano23} and \citet{Boyett24}. We adopt the cosmological parameters from \citet{Planck18}, a flat universe with H$_{0}$ = 67.7 km/s/Mpc and $\Omega_{m}$ =  0.310. With these parameters the universe has an age of 520 Myrs at the redshift of Gz9p3. We use vacuum emission line wavelengths.

\begin{figure*}[!t]
\includegraphics[width=\hsize]{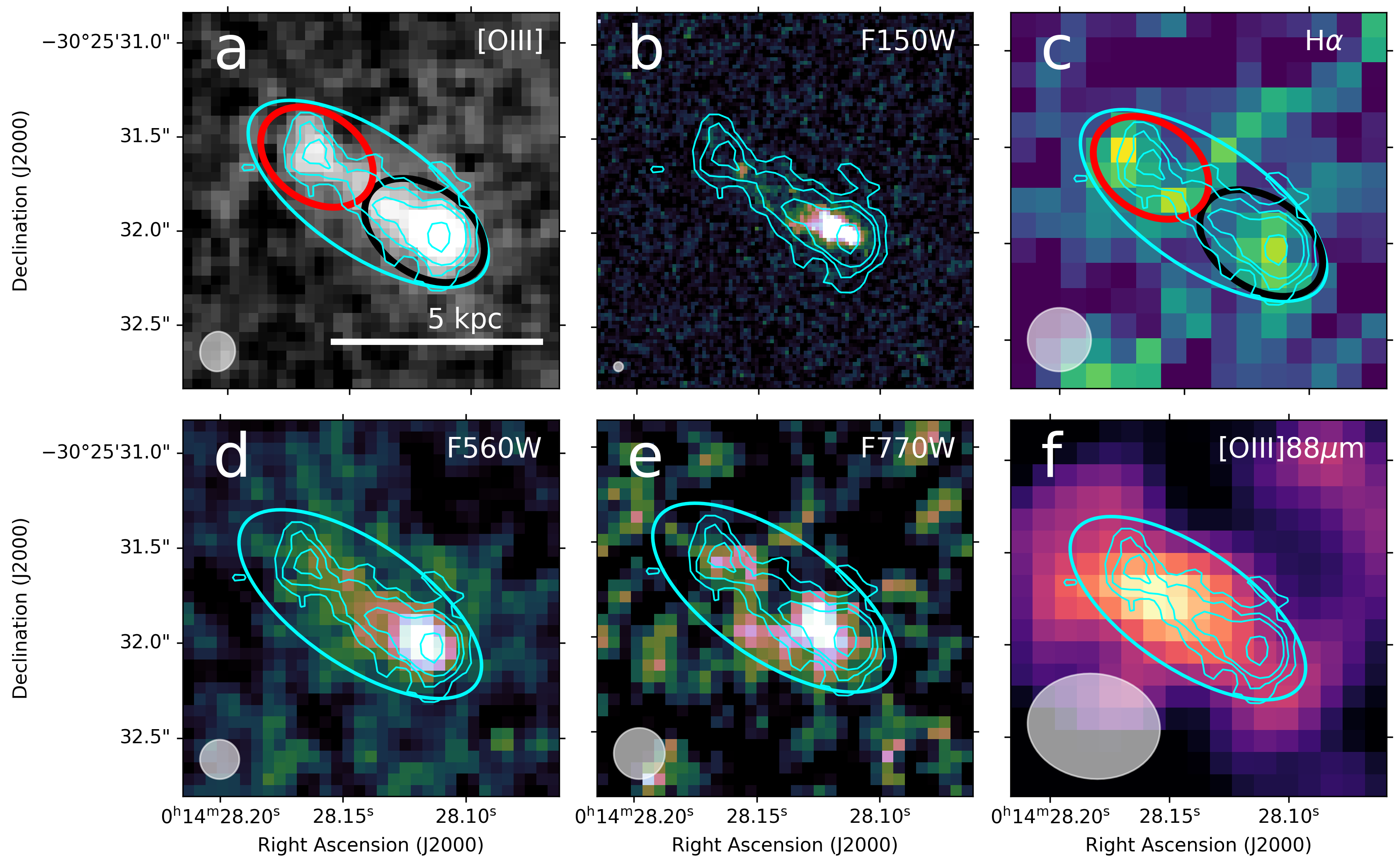}
\caption{Overview of the JWST data of Gz9p3. {\bf a}: NIRSpec IFU continuum subtracted [OIII] emission line map; {\bf b}: NIRCam F150W image \citep{Boyett24} tracing the UV continuum at 1500 \AA, {\bf c}: MIRI MRS \ha\ line map (the continuum is not detected).  {\bf d}: F560W and {\bf e}: F770W MIRI images of Gz9p3. {\bf f}: ALMA [OIII] 88\micron\ emission from \citet{Algera25}. Overplotted on the images are the contours 3, 5, 7 and 15 $\sigma$ of the [OIII] 5008\AA\ emission, the extraction aperture for the integrated galaxy spectrum in cyan, and the two apertures centered on the central clump (black) and the tail (red). The distance marker in panel a is corrected by the lensing magnification. In the bottom-left corner of each image the FWHM of the PSF is shown as a white circle or ellipse.}\label{fig:OIIImap}
\end{figure*}

\section{Observations and data reduction}\label{sec:observations}

\subsection{NIRSpec and MIRI observations}

The observations of Gz9p3 were obtained by JWST with the NIRSpec \citep{Boker23} and MIRI \citep{Rieke15,Wright15,Wright23} Integral Field Units and the MIRI imager \citep[MIRIM][]{Bouchet15} as as part of the  European Consortium MIRI Guaranteed Time Observations (proposal ID 4530, PI: L. Colina) during Cycle 3. The NIRSpec-IFU  \citep{Boker22} data were taken on October 21, 2024, the MIRI-MRS \citep{Wells15,argyriou23}  observations were performed at November 15, 2024 and the MIRI F560W and F777W imaging observations on December 16, 2024.

For the NIRSpec-IFU observations, the PRISM was selected, providing  low-resolution spectra between 0.6 and 5.3 \micron\ of Gz9p3. The observations were obtained with the NRSIRS2RAPID readout pattern using a 6 point medium cycling pattern and 32 groups with one integration per exposure. This results in a total integration time of 2888.6 sec. Immediately after the observations of Gz9p3, one LeakCal observation was executed to correct for leakage from the open shutters of the MSA.

The MRS observations use the LONG configuration, covering the 6.53–7.65 \micron\ wavelength interval in channel 1 \citep{argyriou23}, where H$\alpha$ is expected. The data were obtained in the SLOWR1 reading pattern with 20 groups per integration and each exposure consisted of 2 integrations. We used the 4 point dither pattern optimized for point sources in both the negative and positive direction. In total 5 different dither cycles (of 4 exposure each), alternating between the negative and positive direction. This results in 20 different exposures and a total integration time of 19589 sec. During the MRS observations, we obtained simultaneous imaging of an adjacent field with MIRIM in the F560W and F770W filters for the purpose of astrometrically aligning the MRS observations.

MIRI imaging in F560W and F770W of Gz9p3 was taken separately using the FASTR1 readout pattern. A medium cycling dither pattern was used with 7 dither positions for F560W and 9 for F770W. Each dither position was observed with 138 groups (F560W) and 140 groups (F770W), resulting in 2681 and 3497 seconds of total exposure time for F560W and F770W respectively.

\subsection{Auxiliary data}

The Abell 2744 cluster has been observed with NIRCam in three different observing programs \citep{Paris23}; the JWST-GLASS survey \citep{Treu22}, UNCOVER \citep{Bezanson24} and the Director's Discretionary Time (DDT) program 2756 (PI. W. Chen). The galaxy Gz9p3 is located in the field observed by the DDT program. The imaging of this galaxy has been presented in \citet{Boyett24}. We downloaded the NIRCam imaging of Abell 2744 from the DJA archive which were reduced using the {\sc grizli} software \citep{Brammer23} as described in \citet{Valentino23}. The images are drizzled to a pixel scale of 0.04\arcsec. Additionally, we include the ALMA [OIII] 88\micron\ emission line map recently published by  \citet{Algera25} in our analysis.

\subsection{Calibration}
\subsubsection{NIRSpec}\label{sec:nirspecalib}

The NIRSpec IFU data were calibrated using v1.17.1 of the JWST pipeline \citep{Bushouse23} with CRDS context 1312. Stage 1 of the pipeline was run using standard settings. We applied the snowball flagging in the jump step,  where the {\tt expand\_factor} parameter was set to two, providing the best correction. After stage one we corrected each frame for potential differences in the zero-level following \citet{Perna23} and calculated the median of an empty area on the detector and subtracted those values for the rate file. 
Stage 2 of the pipeline was run with the {\tt nsclean} step to correct for the 1/f noise on the detector. The {\tt nsclean} step provided a better correction for the observed striping than the {\tt clean\_flicker\_noise} step in Stage 1. Inspection of the LeakCal frame revealed emission from several sources where emission leaked through the closed MSA shutters. Five of those sources partly overlapped with the region of the nrs1 detector where the IFU PRISM spectra are recorded. We mask these 5 areas from each of the cal files before processing the cal files further.

%the MSA flagging applied to remove possible contamination of open shutters in the MSA. We used 

We apply several steps to mask the hot pixels in the cal files. First we mask all the very bright pixels. The brightest line in the spectrum of Gz9p3 in the observed wavelength range is [OIII] at 5008 \AA\ (see section \ref{sec:results}). We identify the emission line on the cal files and measure the flux in the brightest pixel in [OIII] line. After that we mask all the pixels with a brighter flux as hot pixel. None of those pixels is close to the [OIII] lines, thus there is no risk for confusing real [OIII] emission for a hot pixel. 
As second step we create a bad pixel mask for each file by subtracting the median of the 6 cal files to each image and employ a sigma clipping in the spectral region bluewards of the [OIII] emission line to iteratively mask 5$\sigma$ outliers from the data. We subtract the continuum of these images to make it easier to find the bad pixels.
The identified bad pixels were added to the dq layer of each of the original cal files. This results in the identification of several additional bad pixels and significantly improves the final spectral cube.
 Finally, stage 3 of the pipeline was run with an outlier detection kernel of 3 x 3 pixels. We created a datacube with a pixel scale of 0.05 \arcsec per pixel using the \emph{drizzle} scheme  \citep{Law23}. 

The angular size of Gz9p3 is smaller than the field of view of the NIRSpec-IFU datacube. We created a sky spectrum by masking out the galaxy and the edges of the cube and taking the median of the remaining pixels (40 \% of the total FOV). This background spectrum is subtracted from each pixel to create the sky subtracted cubes used in the remaining part of the paper.

Finally, we PSF match the NIRSpec following formula 3 and 4 of \citet{Deugenio24}. As reference we use the PSF at the observed wavelength of the [OIII] 5008\AA\ line (5.1675\micron). We convolve each plane in the cube with the quadratic difference between that PSF and the actual PSF of that wavelength. This ensures that when extracting spectra in small apertures we probe the same physical scales at all wavelengths.

\begin{figure*}[!t]
\includegraphics[width=\hsize]{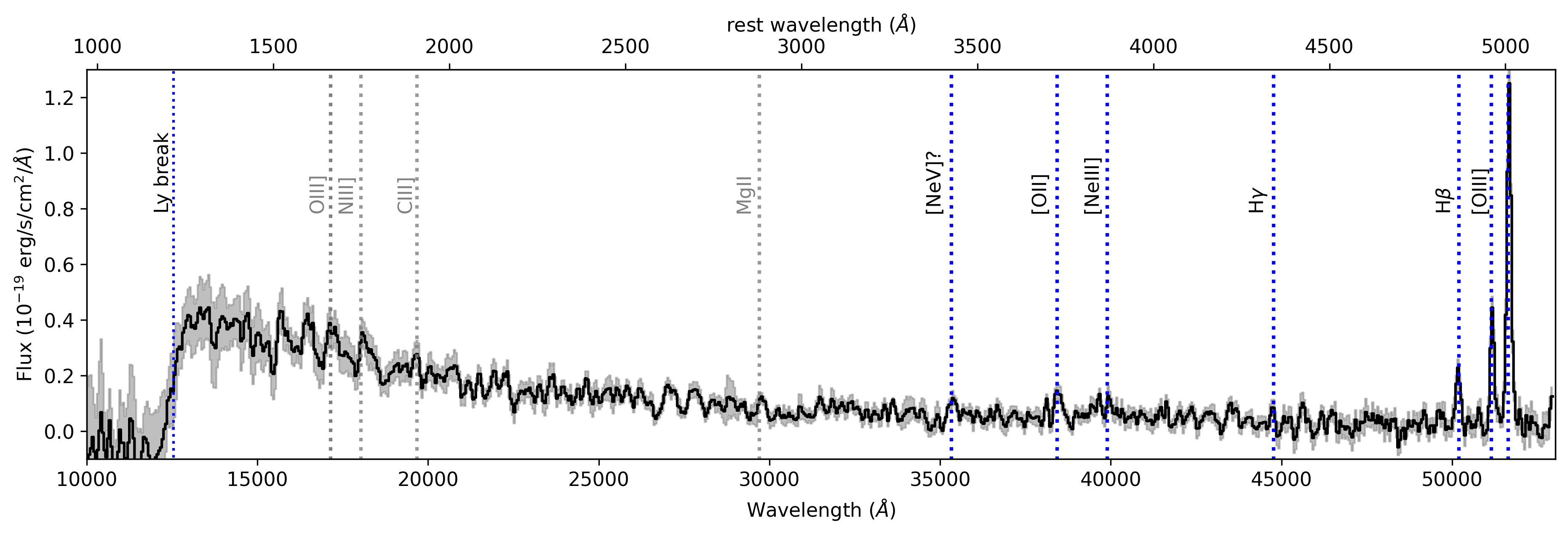}
\caption{Integrated NIRSpec spectrum of Gz9p3, extracted over the elliptical aperture shown in Fig. \ref{fig:OIIImap}. The detected emission lines are marked with blue vertical lines, while the UV lines tentatively detected are marked with gray lines adopting z=9.3127.}\label{fig:NIRSpec_integrated}
\end{figure*}

\subsubsection{MIRI}
The MIRI MRS data were calibrated using v1.20.0 of the JWST pipeline \citep{Bushouse25} with CRDS context 1462. We follow the standard data reduction procedure with some extra steps to improve the data quality following \citet{Alvarez23,Alvarez24}, summarized briefly below. In stage 1 of the pipeline the cosmic showers are flagged in the jump step using a {\tt extend\_snr\_threshold} of 1.4, providing the best cosmic ray flagging without flagging other pixels. We apply a 1/f like correction implemented by \citet{Alvarez24} and \citet{Perna23} due to the vertical striping present in the MIRI detectors. Stage 2 of the pipeline was run using the {\tt clean\_showers} in the straylight step to remove cosmic ray showers not corrected for in the jump step of stage 1. After stage 2 we constructed a background frame by median averaging the cal frames and subtracting the median from each of the cal files. Finally, the background subtracted cube was constructed using the \textit{drizzle} \citep{Law23} algorithm. %As final step we employed the algorithm developed by \citet{Spilker23} to remove the last residual striping due to uncorrected CR showers \citep{Bik24}.

Both the MIRI simultaneous images as well as the F560W and F770W images targeting Gz9p3 were calibrated with  v1.18.1 of the JWST pipeline \citep{Bushouse23} with CRDS context 1364 following the standard data reduction steps for MIRI imaging. We applied additional steps to correct for background gradients and striping artifacts as described in \citet{Alvarez23} and \citet{PerezGonzalez24}. We created one single F560W and F770W image out of the simultaneous imaging and the G9zp3 imaging. In stage 3 we did not align the images to GAIA as the purpose of these images is to calibrate the astrometry of the MRS observations.

\subsubsection{Astrometry}

We aligned all the data used in this paper to the F444W NIRCam image downloaded from the DJA archive. Inspection of the F444W image in the vicinity of Gz9p3 reveals an offset of $\sim$1\arcsec with respect to GAIADR3 \citep{GaiaDR3}. We align the F560W and F770W MIRI images to the F444W NIRCam image by identifying galaxies in both images and calculate the shift between the MIRI and F444W images. The measured offset in the F560W image is applied to the WCS of the MRS datacube.
To align the NIRSpec observations we created a NIRSpec cube with a pixel scale of 0.04\arcsec, to match with the NIRCam F444W image. Then we convolve the NIRSpec cube with the response curve of the F444W filter to construct a F444W image from the NIRSpec IFU data. We aligned the reconstructed F444W to the NIRCam F444W image of Gz9p3 with CROCOA \citep{Runnholm24}, using a 2D cross-correlation to align both images.

\section{Measurements}\label{sec:results}

We detect  spatially resolved emission of Gz9p3 in both the UV to optical continuum and several emission lines such as [OIII], H$\beta$, [OII] in the NIRSpec cube as well as in the H$\alpha$ line in the MRS cube. In the MRS cube, we do not detect the continuum underneath H$\alpha$. We detect Gz9p3 in both the F560W and F770W filter with a clear detection in the F560W image and a fainter detection in F770W. The H$\beta$ and [OIII] lines are included in the F560W filter response, and H$\alpha$ inside the F770W filter. In this section we will present the measurements derived from these datasets.

\subsection{[OIII] and \ha\ emission line maps}\label{sec:linemaps}

We construct an emission line map of the [OIII] emission by adding the 4960\AA\ and 5008\AA\ emission lines. We sum the cube over 5 spectral pixels ($\pm 550$ km/s) for both of the [OIII] lines. We create a map of the [OIII] line emission  by subtracting the underlying continuum determined from the line-free continuum map between 5.19 \micron\ and 5.22 \micron\ (5020 \AA\ and  5120 \AA\ restframe).

We calculate the standard deviation of the background in the emission line map, making sure to mask out the [OIII] emission and the edges of the map. We masked the [OIII] emission by excluding a 0.7 arcsec circle centered on the middle of the emission. We estimate a standard deviation of the background for the [OIII] line map of $1.8\times 10^{-20}$ erg~s$^{-1}$~cm$^{2}$ per pixel. The H$\alpha$ emission line map is calculated by summing over 10 spectral pixels ($\pm$ 250 km/s) between 6.764 and 6.772 \micron. As no continuum is detected underneath H$\alpha$ we do not subtract a continuum image from the resulting H$\alpha$ line map.

Figure \ref{fig:OIIImap} shows the resulting emission line maps together with the F150W image presented by \citet{Boyett24}, the \ha\ line map and the MIRI images as well as the ALMA [OIII]88\micron\ map from \citet{Algera25}. Figure \ref{fig:OIIImap}a shows the [OIII] emission line map, panel b shows the F150W image of Gzp3; both images have the 3, 5, 7 and 15 $\sigma$ contours of the  [OIII] emission overlayed. The contours are calculated using the derived standard deviation of the background.

The [OIII] emission covers the entire galaxy as observed in the continuum and shows an equally clumpy nature. The brightest [OIII] emission is observed towards the two UV bright clumps in the central region of the galaxy. As can be seen at the highest contour in  panel b of Fig \ref{fig:OIIImap}, the [OIII] emission peaks on the fainter of the two UV clumps, suggesting that this clump is the youngest of the two. This core also shows the bluest $\beta_{UV}$ slope  ($\sim$-2.5) in the spatially resolved analysis of \citet{Boyett24}. The faint tail of the galaxy seen in the F150W images shows bright [OIII] emission, while showing little continuum emission.

Panel c of the figure shows the H$\alpha$ linemap as observed with MIRI/MRS. Due to the lower sensitivity of the MRS observations, this line is detected at a lower signal-to-noise (S/N) ratio than the [OIII] emission. We identify three distinct clumps in the emission line map, consistent with the brighter [OIII] emission regions.

\subsection{Integrated spectra of Gz9p3}\label{sec:integrated}

We design extraction  apertures which we use for the extraction of both the NIRSpec and MIRI spectra as well as the MIRI photometry. To obtain the integrated spectrum of Gz9p3 we construct an elliptical aperture with a semi-major axis of 0.74\arcsec (2.6 kpc), a semi-minor axis of 0.325\arcsec\ (1.12 kpc), with the physical distances corrected for the magnification factor, and an angle of 28$^{\circ}$, encompassing the entire galaxy as shown in Fig. \ref{fig:OIIImap} overplotted on the different datasets. 

We use \texttt{photutils} \citep{Bradley24} to extract the spectra inside the  apertures. The errors on the spectra are calculated by median averaging the spectrum of 11 sky apertures, equivalent to the science apertures, placed randomly in the field of view. 

%The resulting NIRSpec spectra of the 5 regions are shown in Fig. \ref{fig:clumpspectra}.  
%We measure the emission line fluxes in the same way as described in Sect. \ref{sec:results} and the values are reported in Tab. \ref{tab:resolved_fluxes}.

We extract the NIRSpec spectrum from the PSF matched data cube. The resulting integrated spectrum is plotted in Fig. \ref{fig:NIRSpec_integrated}. Apart from the two [OIII] emission lines (9$\sigma$ for the 4960\AA, 24$\sigma$ for the 5008\AA\ line) we detect H$\beta$ (6$\sigma$), H$\gamma$ (2$\sigma$) and  [OII] (5 $\sigma$) in the optical restframe. Additionally we tentatively detect [NeIII] and [NeV]. A further analyis outlined below will show that the [NeV] detection is very likely not real. The rest frame UV spectrum shows a lot of artifacts, their strength changes with the choice of algorithm used to create the datacube. This makes it difficult to judge whether a peak in the spectrum is real or not. For completeness we mark the  the UV lines corresponding to peaks seen in the spectrum in Fig. \ref{fig:NIRSpec_integrated} with a gray vertical line. We find peaks in the spectrum corresponding with OIII] at 1663\AA, NIII] at 1749\AA, CIII] at 1908\AA\ and MgII at 2880\AA. However these peaks are not stronger than any of the other artifacts, therefore we choose not to claim the detection of these lines.

In order to check the flux calibration of the NIRSpec spectrum we convolved the NIRSpec cube with the F444W response curve and created a synthetic F444W image. We extracted the flux in the aperture defined above and compared it with the flux extracted from the NIRCam F444W image in the same aperture. We find a flux difference of $\sim$5\%, showing that the flux calibration of NIRSpec is correct and we do not apply any flux scaling to the spectrum.

\begin{table*}
\caption{Measurements of Gz9p3.}  \label{tab:linefluxes}
\centering
\begin{tabular}{rrrr}      % centered columns (4 columns)
\hline\hline
Line flux\tablefootmark{a} &  (10$^{-19} erg~s^{-1}~cm^{-2}$) \\
\hline
& Full galaxy & Central clump & Tail\\
\hline
\hline
{[OII]} $\lambda3727,3729$ & 23.1 $\pm$ 5 & 12.7 $\pm$ 2.2 & $<$7.6\\
{[NeV]} $\lambda3426$  &    $<$21   $\pm$ 13 & $<$13.2  & $<$ 3.0 \\
{[NeIII]} $\lambda3867$  &    6   $\pm$ 4 & 3.1 $\pm$ 1.9 & < 3.0 \\
H$\gamma$ $\lambda4340$  &   8    $\pm$ 4 & 4.3 $\pm$ 1.1  & 3.0 $\pm$ 1.4\\
H$\beta$  $\lambda4861$ & 32 $\pm$ 5 & 11.2 $\pm$ 2.8 & 13.7 $\pm$ 2.5\\
{[OIII]} $\lambda4960$         & 55 $\pm$ 6 & 31.4 $\pm$ 3.6 & 16.0 $\pm$ 2.4\\
{[OIII]} $\lambda5008$         & 189 $\pm$ 8 & 105.2 $\pm$ 3.7 & 57.4 $\pm$ 3.9 \\
H$\alpha$ $\lambda6563$ & 76 $\pm$ 10 & 35 $\pm$ 7.1 & 36 $\pm$ 7.1\\
\hline
\hline
Equivalent widths& ($\AA$)  \\
\hline
EW(H$\beta$)$_{rest}$ & 141 $\pm$ 17 & 53 $\pm$ 11 & 329 $\pm$ 50\\
EW([OIII])$\lambda5008_{rest}$ & 1494 $\pm$ 49 & 731 $\pm$ 19 & 1303 $\pm$ 64\\
EW(\ha)$_{rest}$ & 986 $\pm$ 266 & 695 $\pm$ 280 & $<$ 5000 \\
\hline
\hline
Photometry\tablefootmark{a}& ($\mu$Jy)  \\
\hline
F560W  & 0.33 $\pm$ 0.02 & 0.20 $\pm$ 0.02 & 0.08 $\pm$ 0.01\\
F770W  & 0.22 $\pm$ 0.03 & 0.13 $\pm$ 0.03 & 0.05 $\pm$ 0.02\\
\hline
\hline
Line ratios&  \\
\hline
H$\alpha$/H$\beta$& 2.4 $\pm$ 0.5 &  3.1 $\pm$ 1.0 & 2.6 $\pm$ 0.7\\
%A$_{V,H\alpha}$ & -0.5 $\pm$ 0.5  & 0.2 $\pm$ 0.8 & -0.2 $\pm$ 0.8               \\
H$\gamma$/H$\beta$& 0.3 $\pm$ 0.1 &  0.4 $\pm$ 0.1 & 0.2 $\pm$ 0.1\\
%A$_{V,H\gamma}$ & 3.6 $\pm$ 3.0 & 1.2 $\pm$ 2.0 & 4.4 $\pm$ 2.9 \\

R2 & -0.14  $\pm$ 0.11 & 0.05 $\pm$ 0.13 & $<$ -0.3 \\
O32 & 1.02  $\pm$ 0.09 & 1.02 $\pm$ 0.13 & $>$ 1.0 \\
R23 & 0.93  $\pm$ 0.07 & 1.25 $\pm$ 0.10 &  $<$ 0.8\\
Ne3O2 & -0.58 $\pm$ 0.30 & -0.61 $\pm$ 0.27 & ---\\
R3 & 0.77  $\pm$  0.07 & 0.97 $\pm$ 0.11 & 0.62 $\pm$ 0.08\\
$\hat{R}$ &0.66 $\pm$  0.08 & 0.9 $\pm$ 0.1 & 0.4 $\pm$ 0.2\\
log([OIII]$_{5008\AA}$/[OIII]$_{88\mu m}$) & 0.7 $\pm$ 0.1& 0.8 $\pm$ 0.1 & 0.5 $\pm 0.1$\\
\hline
\hline
Derived properties&  \\
\hline
$\beta_{UV}$ & -2.0 $\pm$ 0.1  & -2.12 $\pm$ 0.05 & -1.6 $\pm$ 0.3\\
F(4200)/F(3500) & 1.1 $\pm$ 0.1 & 1.12 $\pm$ 0.07 & 0.44 $\pm$ 0.3\\ 
SFR$_{H\alpha}$(M$_{\odot}$ yr$^{-1}$) & 13.4 $\pm$ 1.8 & 6.1 $\pm$ 1.3 & 6.5 $\pm$ 1.3  \\
SFR$_{UV}$(M$_{\odot}$ yr$^{-1}$) & 15.1 $\pm$ 0.3 & 12.6 $\pm$ 0.1& 1.3 $\pm$ 0.1\\
%\xi_{ion}$ (erg$^{-1}$ Hz) & 25.29 $\pm$ 0.07 \\
$\xi_{\mathrm{ion, NIRSpec}}$ (erg$^{-1}$ Hz) & 25.4 $\pm$ 0.1 & 25.0 $\pm$ 0.1 & 26.1 $\pm$ 0.2 \\
log(U) & -2.0 $\pm$ 0.1 & -2.0 $\pm$ 0.06 & $>$-2.0 $\pm$ 0.4 \\
Metallicity (12+log(O/H)) & 7.84 $\pm$ 0.05 & 7.95 $\pm$ 0.04 & 7.43$\pm$ 0.06  \\
Burstiness (SFR$_{\mathrm{10 Myr}}$/SFR$_{\mathrm{100 Myr}}$) & 0.9 $\pm$ 0.1 & 0.5 $\pm$ 0.2 & 5 $\pm$ 0.2\\
\hline
\end{tabular}
\tablefoot{All properties are measured in the same aperture, described in Sect. \ref{sec:integrated}.}
\tablefoottext{a}{Observed values, uncorrected for magnification, the MIRI values are corrected for aperture effects.}
\end{table*}

We measure the flux of the detected emission lines using \emph{specutils} \citep{specutils} by fitting a Gaussian profile to each of the emission lines. Before fitting the emission lines the continuum was fitted with a 3rd order Chebyshev polynomial, excluding the region bluewards of the Lyman Break and strong emission lines. Subsequently, we subtracted the continuum from the spectrum. We added a constant in the Gaussian fitting to correct for any residual continuum present.  The errors were calculated using a Monte Carlo simulation by creating 1000 realizations of the spectrum taking into account the error spectrum. We measure the Equivalent Width (EW) of H$\beta$ and [OIII] by normalizing the spectrum by the fitted continuum and apply the same fitting procedure as for the line fluxes. 
The resulting values  are presented in Table \ref{tab:linefluxes}. 

%We find a restframe EW(H$\beta$) = 141 $\pm$ 17 \AA. and the restframe EW([OIII]$\lambda5007$) is 1494 $\pm$ 49 \AA\,(Tab. \ref{tab:linefluxes}). 

\begin{figure}
\includegraphics[width=\hsize]{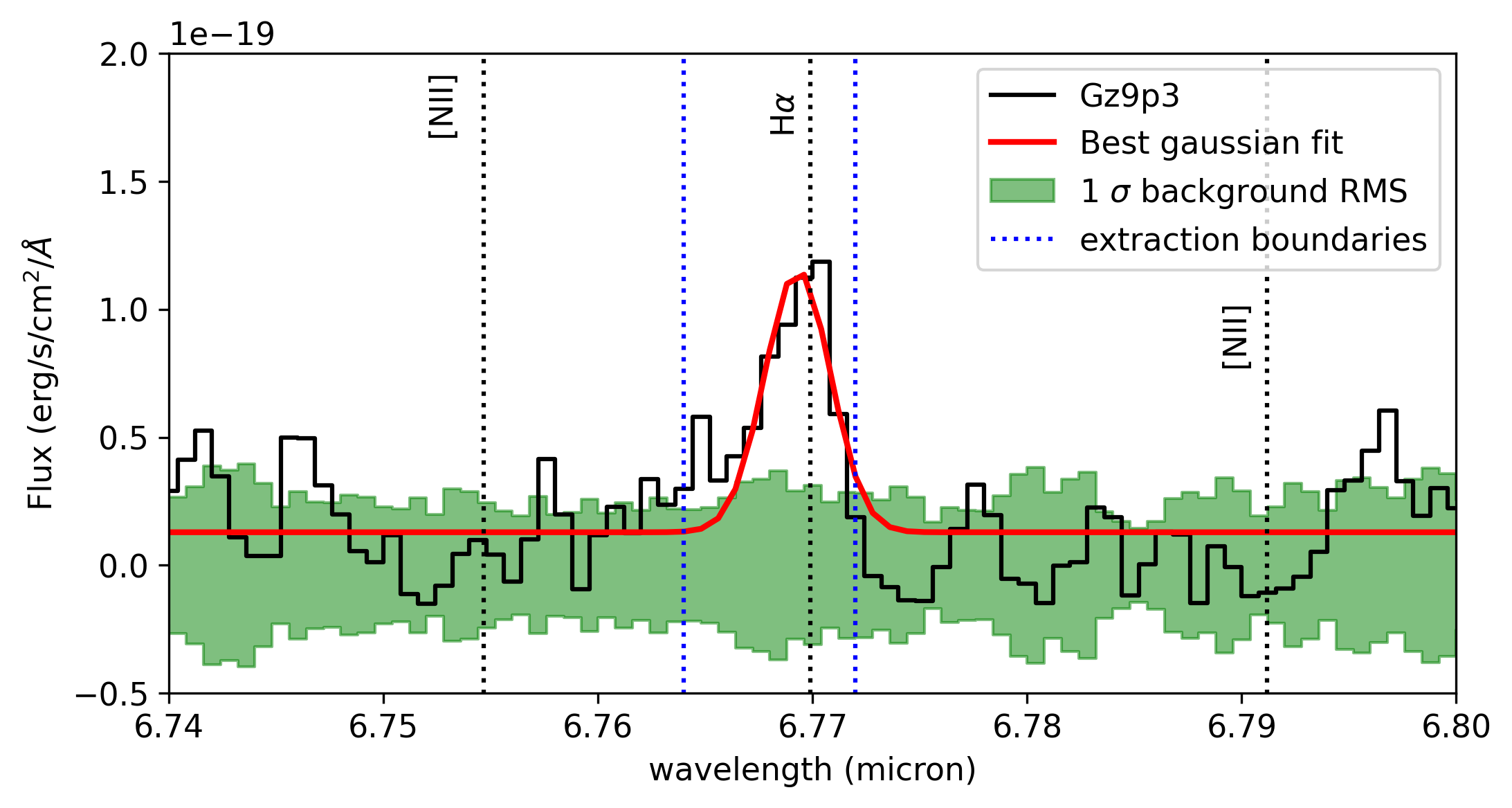}
\caption{MIRI/MRS H$\alpha$ spectrum of Gz9p3 integrated over the elliptic aperture shown in Fig. \ref{fig:OIIImap} with overlayed the best single gaussian fit (see text). \label{fig:MRS_integrated}}
\end{figure}

\begin{figure*}
\includegraphics[width=\hsize]{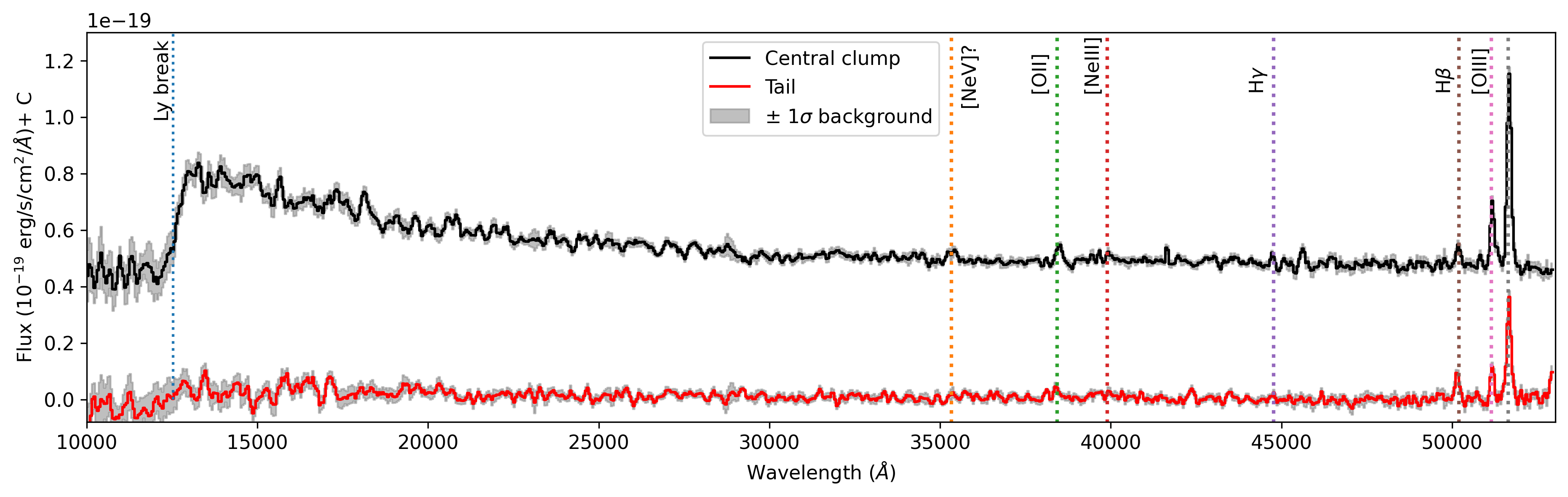}
\caption{NIRSpec spectra of the central clump (black) and the tail region (red) of Gz9p3. The spectrum of the central clump is shifted for clarity.\label{fig:NIRSpec_apertures}}
\end{figure*}

%\subsubsection{MRS spectrum}

The extracted MRS spectrum is plotted in Fig. \ref{fig:MRS_integrated}.  The error on the MRS spectrum is estimated by the standard deviation of the spectra extracted from 14 apertures randomly placed in the MRS field of view and plotted as the green histogram in Fig. \ref{fig:MRS_integrated}. We do not detect the [NII] lines at 6548.05 \AA\ and 6583.45\AA.

We calculate the line flux of H$\alpha$ by integrating the line profile between the two dotted blue lines in Fig. \ref{fig:MRS_integrated} and detect the H$\alpha$ line with a significance of 7.5 $\sigma$. We estimate the aperture correction for our elliptical aperture by simulating 3 point sources mimicking the H$\alpha$ distribution using \texttt{galfit} \citep{Peng10}. We calculate the scene using the MRS PSF models \citep{argyriou23} and we derive an aperture correction of 1.35. The final, aperture corrected flux is given in Tab. \ref{tab:linefluxes}.

The H$\alpha$ emission line is spectrally resolved and shows an indication for an asymmetric profile. We fit a single Gaussian to the observed profile and measure a systemic velocity consistent with 0 km s$^{-1}$ (see Sect. \ref{sec:redshift}) and  a FWHM of 138 $\pm$ 49 km s$^{-1}$ , corrected for the instrumental broadening \citep{Labiano21,Jones23}. The best fitting Gaussian is plotted in Fig. \ref{fig:MRS_integrated} in red. The error on the FWHM is determined by the Monte Carlo approach as described above. Fitting a double gaussian to the MRS spectrum reveals a fainter, blue shifted component detected at 2 $\sigma$ level. 
 %T

\subsection{MIRI photometry}

The MIRI F560W and F770W images of Gz9p3 (Fig. \ref{fig:OIIImap}) reveal a detection in both bands. The morphology of the galaxy in the two bands is different. The peak of the F560W emission towards the central clump is coinciding with the peak of the [OIII] emission, while the peak of the F770W is shifted and co-located with the brightest continuum clump in e.g. the F150W image. The F560W filter contains both H$\beta$ and the [OIII] lines in the filter throughput, where especially the bright [OIII] could affect the appearance of the galaxy in this filter. The F770W filter contains the intrinsically fainter H$\alpha$ line and the galaxy is detected at lower S/N and reveals mainly the bright continuum clumps.

We measure the brightness of the galaxy on 10\arcsec$\times$10\arcsec\ cutouts of the calibrated images. In order to remove any background residual, we construct a background using the task \texttt{Background2D} in photutils \citep{Bradley24} after masking out galaxies in the FoV. We removed a small background gradient from the data.
%We smoothed the background image with a gaussian kernel with a width of 3 pixels to create a segmentation map with a threshold of 1.5 $\times$ the rms of the unconvolved image. This resulted in the detection of Gz9p3 as a single galaxy in both the images with the highes
%As the galaxy is detected with the highest S/N we use the F560W image to calculate the kron aperture. This kron aperture is larger than the aperture we use for the spectroscopy. As we will use the photometry to derive the \ha\ EW, we only use the central position of the kron aperture and the size and orientation of the elliptical aperture used for spectroscopy and perform photometry using this aperture on the unconvolved images in both filters.

%The kron aperture is an elliptical aperture with a semi major axis of 16 (1\arcsec) and a semi minor axis of 8 pixels (0.55\arcsec) and is shown in Fig. \ref{fig:OIIImap}. 
 We perform photometry using the spectroscopic aperture as displayed in Fig. \ref{fig:OIIImap}. We calculate the photometric errors by randomly placing 12 apertures of the same size in the 10\arcsec$\times$10\arcsec\ cutout and use the standard deviation of the flux in these apertures as error. We measure a flux of 0.33 $\pm$ 0.02 $\mu Jy$ in F560W and 0.22 $\pm$ 0.03 $\mu Jy$ in F770W. %Consistent with the NIRCam photometry presented in \citet{Boyett24}. 
%{\bf do we want an SED fit?}

\subsection{Spatially resolved spectra}

In order to look for spatially resolved variation of the spectral properties we construct two sets of smaller apertures. First, following \citet{Boyett24} we divide the galaxy in the central clump region and the tail region. The two elliptical apertures covering the central clump region and the tail region  are shown as black and red  apertures in Fig. \ref{fig:OIIImap}. The apertures have a major axis of 0.35 (0.32)\arcsec\ for the central clump (tail) aperture, a minor axis of 0.23\arcsec\ and an angle of 28$^{\circ}$ for both apertures. 

\begin{figure}
\includegraphics[width=\hsize]{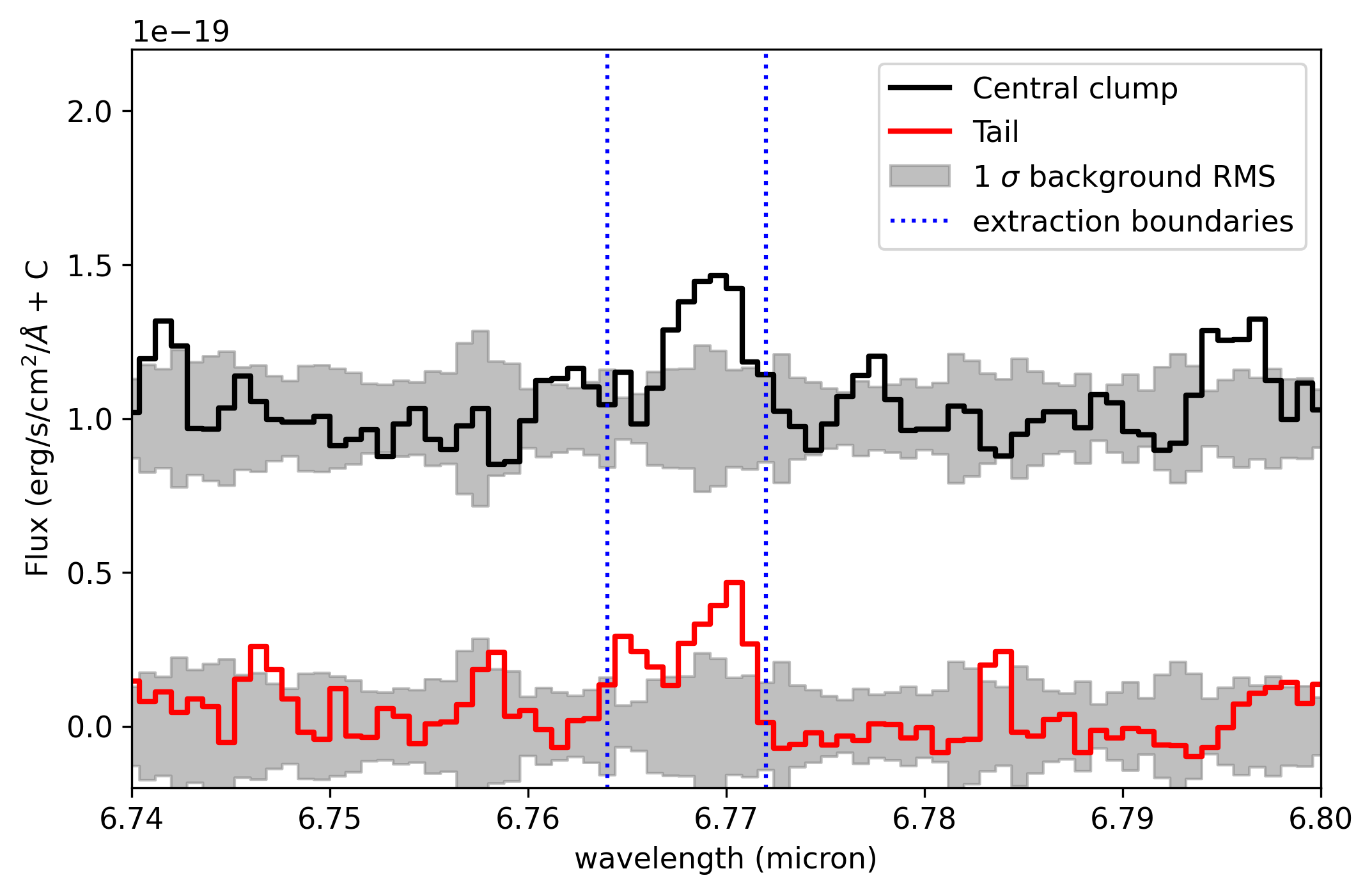}
\caption{MIRI/MRS H$\alpha$ spectra of the central clump (black) and the tail region (red) of  Gz9p3.  The spectrum of the central clump is shifted for clarity. \label{fig:MRS_apertures}}
\end{figure}

The extracted NIRSpec and MIRI-MRS spectra in the two smaller apertures focused on the central clump and the tail of the galaxy are shown in Figs. \ref{fig:NIRSpec_apertures} and \ref{fig:MRS_apertures}. The NIRSpec spectra look very different, where the spectrum of the central clump shows a strong continuum with a clear Lyman break as well as strong line emission. The spectrum of the tail region barely shows continuum emission and is dominated by line emission of predominantly [OIII] and H$\beta$. This is consistent with the morphology seen in the NIRCam imaging \citep[Fig. \ref{fig:OIIImap},][]{Boyett24}, where only a few faint clumps are detected towards the tail, while [OIII] is detected with a much higher significance. Both apertures show significant H$\alpha$ emission in the MRS spectra. The spectrum of the tail region reveals the faint blue shifted emission seen in the integrated spectrum, suggesting a velocity gradient between the clumps.

\begin{figure}
\includegraphics[width=\hsize]{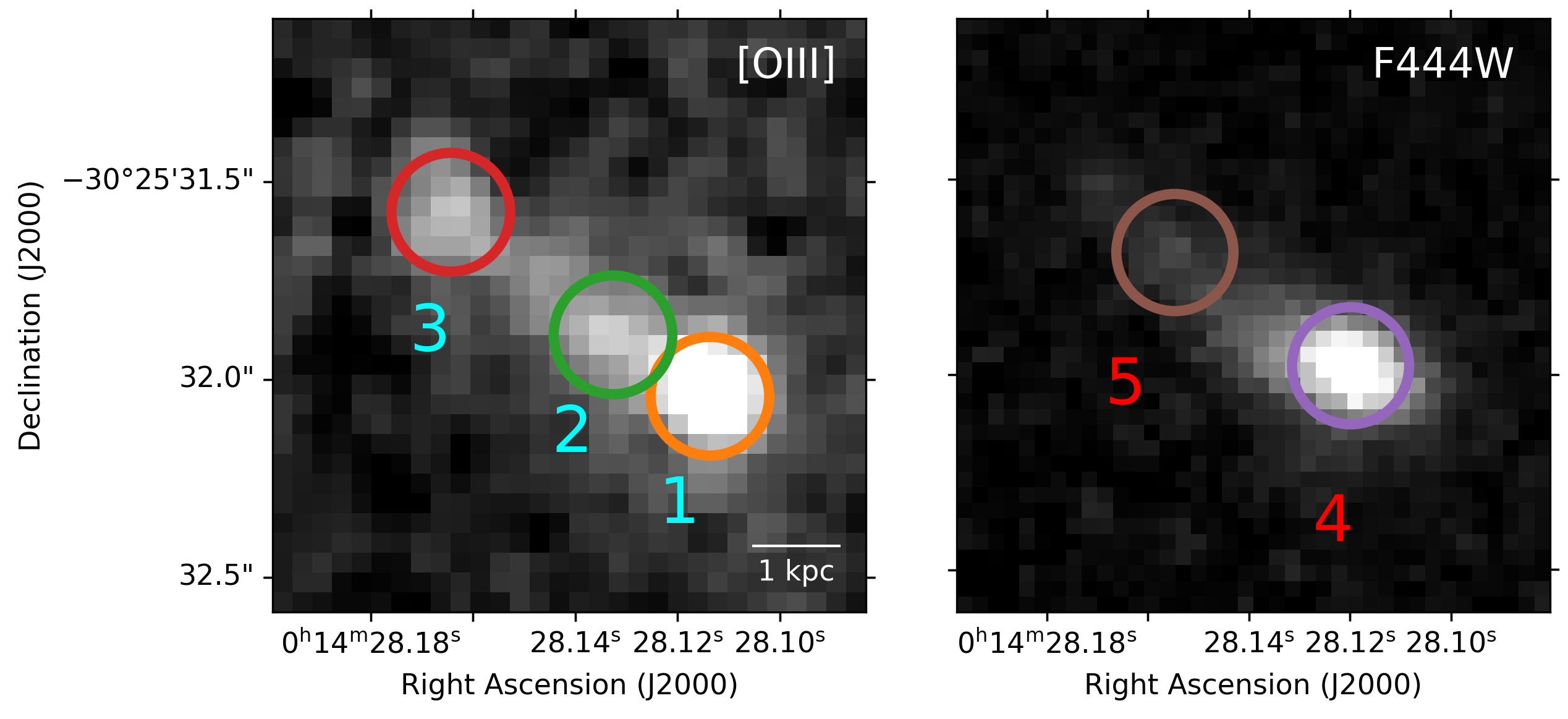}
\caption{[OIII] line map (left) and F444W image (right) of Gz9p3 with overlayed the apertures used to extract the spectra of the spatially resolved analysis. \label{fig:apertures}}
\end{figure}

Due to the relatively high SNR of the NIRSpec data, we define a set of smaller apertures for the NIRSpec data only. 
We use the [OIII] map and the NIRCam imaging of Gz9p3 for guidance to select five smaller circular apertures with a radius of 3 pixels (0.15\arcsec), resulting in an aperture larger than the FWHM of the spectrograph. Based on the [OIII] map we select 3 apertures covering the 3 clumps (apertures 1 -- 3, Fig. \ref{fig:apertures}, left panel). Additionally, we select two apertures based on the F444W image (Fig. \ref{fig:apertures}, right panel), where one aperture is placed on the brightest continuum peak \citep[similar to the inner aperture in][]{Boyett24} and one on the faint continuum clump where [OIII] appears to be faint \citep[corresponding to the aperture on the tail,][]{Boyett24}. Note that aperture 1 and 4 partly overlap, due to the limited spatial resolution of NIRSpec at 5 \micron. Clumps 1 and 4 correspond to the central region of the galaxy, likely associated with the nuclei of the proto-galaxies involved in the merger. Clumps 3 and 5 are clearly associated with the tail of Gz9p3, while clump 2 is could be associated to the inner regions of the tail, emitting mostly line emission. The extracted spectra of the five apertures are shown in Fig. \ref{fig:clumpspectra}.

\subsection{Redshift}\label{sec:redshift}

The high spectral resolution NIRSpec-MSA observations presented by   \citet{Boyett24} resulted in a spectroscopic redshift determination of $z$ = 9.3127 $\pm$ 0.0002. A slightly higher redshift ($z = 9.325 \pm 0.001$) was found based on low spectral resolution NIRSpec PRISM observations by \citet{Fujimoto24}. We measure the redshift of Gz9p3 on the integrated MIRI H$\alpha$ spectrum (Fig. \ref{fig:MRS_integrated}) as that has the highest spectral resolution. From the fitting of the \ha\ line we find a redshift of $z$ = 9.3118 $\pm$ 0.0009, consistent with the redshift derived by \citet{Boyett24}. Our error on the redshift is larger due to the weak H$\alpha$ detection, therefore we  will use $z$ = 9.3127 $\pm$ 0.0002 as derived by \citet{Boyett24} for the remainder of this paper.

%We measure the redshift of the galaxy in the integrated spectrum covering all the [OIII] emission. We fit a gaussian profile to the  brightest [OIII] emission line and find a fitted redshift of z = 9.31730 $\pm$ 0.00036.

\section{Physical properties}\label{sec:physical_properties}

In this section we use the measurements presented in the previous section to derive physical properties of the galaxy. We reassess the metallicity of Gz9p3, derive properties related to the stellar content of Gz9p3, the star formation rate, output of ionizing photons, UV slope and search for the presence of a Balmer break. Finally we will plot Gz9p3 in emission line ratio diagrams to compare it to other low- and high redshift galaxies. We will make use of the following diagnostic ratios:

\[R2 = \log(\frac{[OII]\lambda3727,3730}{H\beta})\]
\[R3 = \log(\frac{[OIII]\lambda5008}{H\beta})\] 
\[R23 =\log(\frac{[O III]\lambda4960,5008 + [O II]\lambda3727,3730}{H\beta})\]  
\[O32 =\log(\frac{[O III]\lambda4960,5008}{[O II]\lambda3727,3730})\]
\[Ne3O2 =\log(\frac{[NeIII]\lambda3869}{[OII]\lambda3727,3729})\]
\[\hat{R} = 0.47  \times R2 + 0.88 \times R3\]

\subsection{Extinction}\label{sec:extinction}

We use the observed \ha, \hb\ and H$\gamma$ emission line fluxes to place constraints on the extinction in Gz9p3. We calculate the line ratios in both the integrated spectrum and the spectra of the central clump and the tail. As intrinsic line ratios following the case B approximation \citep{Osterbrockbook} we use \ha\ over \hb\ = 2.85 and H$\gamma$ over \hb\  = 0.469 for a electron temperature of 10,000 K and an electron density of n$_e$ = 1000 cm$^{-3}$ as measured for galaxies at $z \sim 9-10$ \citep{Abdurrouf24}. Based on the [OII] doublet, \citet{Boyett24} measured a  n$_e$ = $590^{+590}_{-250}$ cm$^{-3}$.  We calculate the extinction using the \citet{Gordon03} extinction law and an $R_V$ = 3.1. Table 1 lists the observed \ha\ over \hb\  and H$\gamma$ over \hb\ ratios for the 3 spectra. 

From the integrated spectrum, we measure a H$\alpha$ over H$\beta$ of 2.4 $\pm$ 0.5, consistent within 1 sigma with the theoretical ratio (2.85) of zero extinction. The value of 2.4 is lower than the theoretical value, which results in a (nonphysical) negative extinction. This could be affected by non-perfect flux calibration of either NIRSpec or MIRI. The extinction derived from the H$\gamma$ over \hb\ ratio suggests a much larger value, but is,
due to the low detection significance (2$\sigma$) of H$\gamma$, not significant. 
Similar values are found for the spectrum of the smaller central clump and  tail apertures of Gz9p3, where the extinction derived from \ha\ over \hb\ provides the strongest constraint, suggesting no measurable extinction. This is consistent with the results of the SED fitting of \citet{Boyett24} where also very low or no extinction was found. The  H$\gamma$ over \hb\ ratio has a too large an error to provide meaningful constraints.

%The \ha\ over \hb\ ratio of the central clump spectrum is very high and results in an extinction of A$_V$ = 2.7 $\pm$ 1 magnitudes, where the extinction derived from the H$\gamma$ over \hb\ ratio is a bit lower, but does not put any meaningful constraint due to the error of almost 2 magnitudes. This high value found for the central clump is inconsistent with the shape of the UV continuum of the spectrum and the measured blue $\beta$ slope ($\beta$ = -2.27 $\pm$ 0.07) as well as with the results from \citet{Boyett24}, who found an A$_{V}$ = 0.03$^{+0.04}_{-0.02}$ mag. 

Based on these measurements as well as the extinction derived by \citet{Boyett24} we do not correct the emission lines for extinction, as they are mostly consistent with no extinction and would have introduced large uncertainties due to the faint \ha\ detection.

\subsection{Star formation rate and burstiness}
We calculate the \ha\ luminosity and derive the instantaneous star formation rate (SFR) (under the assumption of negligible extinction, Sect. \ref{sec:extinction}).
 Using the conversion of \citet{Theios19} for 20\% solar metallicity assuming a \citet{Kroupa01} initial mass function (as measured for Gz9p3 in Sect. \ref{sec:metallicity}), we derive a star formation rate of 13.4 $\pm$ 1.8 M$_{\odot}$ yr$^{-1}$. This is rather equally split between the main clump and the tail region (Tab. \ref{tab:linefluxes}). For the clumps extracted from the NIRSpec spectra alone we calculate the SFR from the \hb\ line instead (Tab. \ref{tab:resolved_fluxes}).

We use the UV luminosity measured from our spectra to derive the SFR sensitive to the last 100 Myrs \citep{Kennicutt12,Raiter10}. We calculate the luminosity at 1500 \AA\, (L$_{1500}$) directly from the extracted NIRSpec spectrum by converting it to F$_{\nu}$ and correcting for the distance as well as K-correction. We measure the L$_{1500}$  by fitting a powerlaw to the UV spectrum to avoid that the strong artifacts influence the derived L$_{1500}$.  The error on the L$_{1500}$ is calculated using a Monte Carlo approach as described in Sect. \ref{sec:integrated}. We calculate the SFR$_{UV}$ from the  L$_{1500}$ following \citep{Kennicutt12,Raiter10} converting the values to a \citet{Kroupa01} IMF in order to compare them directly with the measured SFR from \ha. We measure a total  SFR$_{UV}$ = 15.1 $\pm$ 0.3 M$_{\odot}$ yr$^{-1}$ (Tab. \ref{tab:linefluxes}), similar to the SFR averaged over the last 100 Myr  derived from SED fitting by \citet[19$^{+5}_{-6}$  M$_{\odot}$ yr$^{-1}$]{Boyett24}.

The SFR$_{UV}$ for the central clump is 12.6 $\pm$ 0.1  M$_{\odot}$ yr$^{-1}$, showing that most of the UV emission of Gz9p3 originates in the central clump, while the SFR$_{UV}$ of the tail region is only 1.3 M$_{\odot}$ yr$^{-1}$. We can compare these values to the results of the SED fitting, where \citet{Boyett24} perform SED fitting of the photometry and spectra combined only of the central region covered by their MSA slitlets and recover a SFR$_{100Myr}$  of 9.1 $\pm$ 0.6  M$_{\odot}$ yr$^{-1}$. 
Their photometry-only SED fitting of the main component results in a much lower SFR$_{100Myr}$ (2.2 $^{+2.5}_{-0.7}$  M$_{\odot}$ yr$^{-1}$. For the tail region only a photometric SED fit is available with a SFR$_{100Myr}$ of 0.2 $^{+2.5}_{-0.7}$  M$_{\odot}$ yr$^{-1}$, similar to our measurement.

The ratio of the SFR$_{10Myr}$ over the SFR$_{100Myr}$ is used as a measure for the burstiness of star formation \citep{Atek22}, where a ratio above unity is indicative of a bursty phase of star formation. 
Galaxies at low-mass and higher redshift tend to show higher burstiness parameters \citep[e.g.][]{Navarroburst24,Langeroodi24,Looser25}. For G9pz3 we find a burstiness parameter of 0.9 $\pm$ 0.1 for the integrated emission of the galaxy, suggesting that the star-formation is rather constant over the last 100 Myrs. For the central clump and the tail region we find a burstiness parameter of 0.5$\pm$ 0.2 and 5.0 $\pm$ 0.2 respectively.  The low burstiness parameter found in the central region is consistent with the older ages found for this clump \citep[50 - 150 Myrs,][]{Boyett24}, while the tail region is barely detected in the F150W images, while clearly visible in \ha\ as well as other nebular lines ([OIII]), suggesting that this is a very recent site of star formation. 

%For the tail region we find a much higher value; 32.5 $\pm$ 17.5, suggesting that the star formation in the tail is very recent and the stellar population very young.

\begin{figure*}
\includegraphics[width=\hsize]{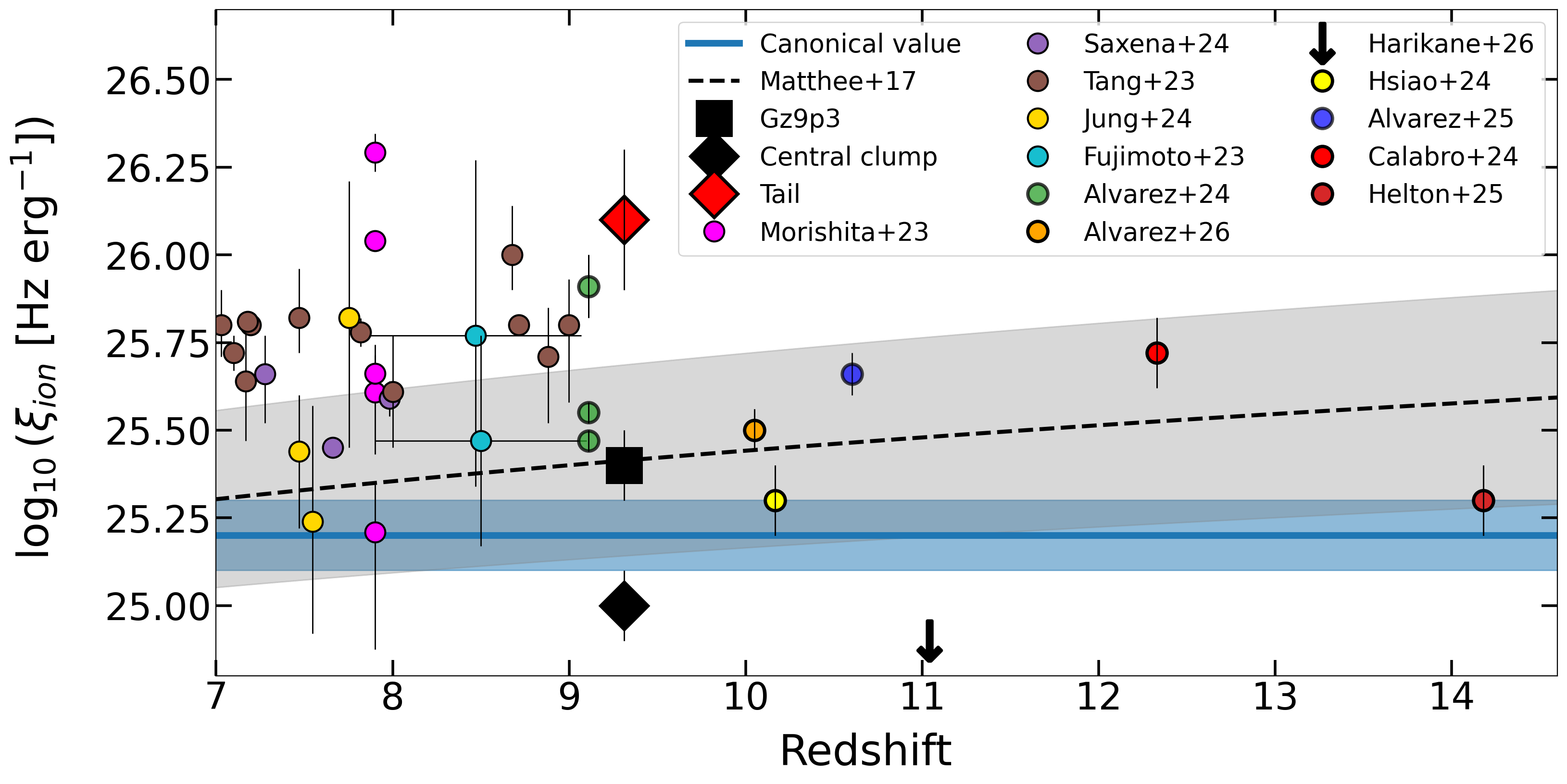}
\caption{Ionizing photon production efficiency ($\xi_{ion}$) as function of redshift. The integrated value for Gz9p3 (black square) and the values for the central clump and the tail of Gz9p3 (black and red diamonds) are compared to spectroscopic samples \citep{Fujimoto23,Tang23,Morishita23,Saxena24,Jung24}.
The green circles show the location of MACS1149-JD, \citep{Alvarez24} The galaxies above redshift 10 shown are GN-z11 \citep[blue circle,][]{Alvarez25},  MACS0647-JD \citep[yellow circle,][]{Hsiao24}, UNCOVER-26185 \citep[orange circle]{Alvarez26}, CEERS2-588 \citep[black arrow]{Harikane26}, GHZ2/GLASS-z12 \citep[red circle,][]{Calabro25} and JADES-GS-z14-0 \citep[dark red circle][]{Helton25}. The canonical value and its uncertainty \citep{Robertson13} is shown as the blue band. The grey band increasing with redshift shows the relation between $\xi_{ion}$ and redshift from \citet{Matthee17}.
\label{fig:xi_ionvsz}}
\end{figure*}

\subsection{Ionizing photon production efficiency}

We calculate the ionizing photon production efficiency ($\xi_{ion}$), the ratio of the production rate of  ionizing photons and the non ionizing UV flux. We calculate the number of ionizing photons from the \ha\ flux, following \citet{Alvarez24} and use the L$_{1500}$ derived in the previous section. We calculate $\xi_{ion}$ from the integrated spectrum as well as the central clump and tail aperture using their \ha\ fluxes (assuming no extinction) and derived L$_{1500}$ from the NIRSpec spectra. Additionally we calculate $\xi_{ion}$ for the 5 NIRSpec only apertures where we use \hb\ for the calculation of the number of ionizing photons. 

For the integrated spectrum of Gz9p3 we derive a value of log $\xi_{ion}$ = 25.41 $\pm$ 0.05 Hz erg$^{-1}$. When we instead derive the $\xi_{ion}$ using the M$_{UV}$ found by \citet{Boyett24} we find a $\xi_{ion}$ = 25.29 $\pm$ 0.07 Hz erg$^{-1}$, the small difference could be caused by the differences in aperture to measure the M$_{UV}$. Figure \ref{fig:xi_ionvsz} shows the derived value of $\xi_{ion}$ compared to spectroscopic measurements of other high redshift galaxies between z= 7 and 13 as well as the canonical value of 25.2 $\pm$ 0.1 Hz erg$^{-1}$ \citep{Robertson13} and the relation between $\xi_{ion}$ and redshift derived by \citet{Matthee17}. The integrated $\xi_{ion}$ value for Gz9p3 is consistent with both the canonical value and the \citet{Matthee17} relation, similar to many high redshift galaxies. When we consider the two  apertures in Gz9p3, we see large differences in the same galaxy. The central clump has a value below the canonical value  ($\xi_{ion}$ =  25.0 $\pm$ 0.1  Hz erg$^{-1}$), consistent with the evolved stellar population found by \citet{Boyett24}. The tail region, on the other hand, has a very high $\xi_{ion}$ ( 26.1 $\pm$ 0.2  Hz erg$^{-1}$). Recent studies of the spatially resolved  $\xi_{ion}$ in high redshift galaxies have revealed similar difference between different regions in the galaxy or different galaxies in interacting/merging systems \citep[e.g.][]{Alvarez24,PrietoJimenez25,Komarova25}.

Finally, we calculate the \ha\ EW by combining the F770W photometry and the \ha\ spectroscopy for both the integrated spectra and the two apertures.
We divide the \ha\ line flux by the effective width of the F770W response curve and calculate the flux of \ha\ in the F770W, from where we calculate the real continuum flux of Gz9p3 in F770W. Dividing the \ha\ line flux by the continuum results in a restframe EW of 986 $\pm$ 266\AA\ (log(EW) = 2.99 $\pm$ 0.12, Tab. \ref{tab:linefluxes}) for the integrated apertures. For the central aperture we find a lower value of 695 $\pm$ 280 \AA, while for the tail aperture we are only able to place an lower limit of 5000 \AA\ due to the very faint continuum in that aperture. These values reflect the differences found in $\xi_{ion}$ where the tail has a very young population currently forming stars while the central clump already contains an older stellar population. 

Several studies find a relation between EW$_{H\alpha}$ and $\xi_{ion}$ \citep[e.g.][]{Prieto23,Rinaldi24} and find strong correlations between the EW(\ha) and $\xi_{ion}$ of galaxies. \citet{Komarova25} derived $\xi_{ion}$ for clumps in the REBELS galaxies at z=6-7 and find variations inside galaxies and recover the trend between EW(\ha) and $\xi_{ion}$ also on smaller spatial scales. Comparing the values for Gz9p3 of $\xi_{ion}$ and EW$_{H\alpha}$ of high redshift galaxies as compiled by \citet{Alvarez24,PrietoJimenez25} we find that the integrated as well as the central clump follow the trend observed between  EW$_{H\alpha}$ and $\xi_{ion}$. The $\xi_{ion}$ value for the tail (26.1 $\pm$ 0.2) is high and consistent with a young star burst \citep{Stanway23,PrietoJimenez25}

\subsection{Metallicity}\label{sec:metallicity}

%The metallicity of Gz9p3 is estimated by \citet{Boyett24} based on their detection of  [NeIII] and [OII]. Using several calibrations between Ne3O2 and metallicity, the authors found a value of 12 + log(O/H) = 7.6 $\pm$ 0.2 with a systematic uncertainty of 0.3. \citet{Scholte25} summarizes the existing calibrations for the metallicity based on the Ne3O2 index and shows a very large spread. Depending on the calibration used, the authors find values ranging from 7.2 to 7.9. The Ne3O2 index measure by \citet{Boyett24} is a lot higher (-0.09) than the one measured in this paper (-0.58). This is caused by a much stronger [OII] detection in our spectra and could be related to the difference in apertures.  

% The NIRSpec-PRISM spectrum contains several other strong lines and can be used as metallicity estimate. Strong line ratios such as the R2, R3 and O32 depend also strongly on the ionization of the gas \citep{Kewley2002ep}. The R23 diagnostic is less sensitive to ionization as it includes both single and double ionized Oxygen. However, this index shows a turnover with metallicity with a plateau around 12 + log(O/H) $\sim$ 8, where it becomes insensitive to metallicity.

The NIRSpec-prism spectrum presented here probes rest-optical high-significance ''strong'' Balmer and Oxygen emission lines which can be used as accurate proxies for the gas-phase metallicity. Here, we estimate the gas-phase metallicity of each region based on its measured [OII]$\lambda\lambda3727,29$; H$\beta$; [OIII]$\lambda\lambda4960,5008$ and H$\alpha$ line fluxes. For this purpose, we adopt the nonparametric probabilistic strong-line calibration presented in \cite{langeroodi+26}, employed through the \texttt{genesis-metallicity} package\footnote{\url{https://github.com/langeroodi/genesis_metallicity}} \citep{2025zndo..15306794L}. Instead of fitting polynomials to the 2-dimensional projections of the strong emission line ratios vs. gas-phase metallicity parameter space, \texttt{genesis-metallicity} estimates the probability density function (PDF) in the full multi-dimensional parameter space directly based on the distribution of the calibration data. This PDF is then used to estimate the gas-phase metallicity for any input combination of strong emission line ratios. By using the EW(H$\beta$) as an additional input parameter for capturing the ionization state, \texttt{genesis-metallicity} achieves a universal redshift-independent calibration. As reported in Table \ref{tab:linefluxes}, we measure $12+\log({\rm O/H})$ values of $7.84 \pm 0.05$, $7.95 \pm 0.04$, and $7.43 \pm 0.06$ for the full galaxy, central clump, and the tail, respectively. This corresponds to a $\sim$ 15\% solar metallicity for Gz9p3, assuming a solar metallicity of 12 + log(O/H) = 8.69 \citep{Asplund09}, 20 \% of the solar value for for the central clump and only 5 \% for the tail region.

For further verification we estimate the gas-phase metallicity of Gz9p3 using polynomial strong-line calibrations. More specifically, following \citet{Scholte25} we use the $\hat{R}$ index, introduced by \citet{Laseter24}, a linear combination of the R2 and R3 index, showing a tighter correlation with metallicity than for example the O32 or Ne3O2 indices. The $\hat{R}$ index measured from the integrated spectrum of Gz9p3 is 0.66 $\pm$ 0.08 (Table \ref{tab:linefluxes}), which would correspond to a  12+log(O/H) = 7.8 $\pm 0.2$ adopting the calibration of \citet{Scholte25}, consistent with the \texttt{genesis-metallicity} results. These values are also consistent with the values derived using the R23 calibration from \citet{Sanders24} and \citet{Scholte25}. \citet{Boyett24} estimated the metallicy from the Ne3O2 ratio, limited by the few emission lines present in their MSA spectrum and found a value of 12 + log(O/H) = 7.6 $\pm$ 0.2 with a systematic uncertainty of 0.3, consistent with our results.
%This value is similar to the value derived from Ne3O2 ratio by  Comparing the R2, R3 and O32 indices to the various calibrations summarized in \citet{Scholte25}, provide similar results with even larger uncertainties.

%In the spectrum of the central clump we find a large value for $\hat{R}$ (0.88 $\pm$ 0.11), above the values expected by the $\hat{R}$ calibration from \citet{Scholte25}. This value is driven by a high R3 value, possibly caused by a non stellar excitation (Sect. \ref{sec:emissionratios}). The metallicity of the tail could be derived (12+log(O/H) = 7.3 $\pm$ 0.3) and is lower than the overall metallicity of Gz9p3, consistent with a merger scenario where two galaxies with different metallicity merge. 

%Table \ref{tab:resolved_fluxes} lists the $\hat{R}$ values for the five clumps and shows a large variation, from $\hat{R}$ = 1.2 $\pm$ 0.2 in clump 2 to $\hat{R}$ =  0.5 $\pm$ 0.1 in clump 5 (the even lower value of clump 3 has a very large error. Clump 1 and 4 show values consistent with the integrated value, however the two clumps in the tail of Gz9p3 (clump 3 and 5)  show a somewhat lower value. For clump 5 we find a metallicity of 12+log(O/H) = 7.5 $\pm$ 0.3, slightly lower and similar to the value of the tail region, but statistically consistent with the integrated metallicity.

\subsection{UV slope and Balmer break.}
The slope of the rest-frame UV continuum parameterized as $\beta$ (f$_{\lambda} \propto \lambda^{\beta}$) is a powerful diagnostics tracing the stellar population. The value of the UV slope is set by the properties of the young stellar population with a contribution of the nebular emission and reddened by possible extinction \citep[e.g][]{Calzetti94,Cullen23}. 

\citet{Calzetti94} calculate the value for $\beta$ by  fitting a power law to the UV part of the spectrum between 1250 and 2600 \AA\ rest wavelengths. We follow this approach and change the blue wavelength to 1400 \AA\ in order to exclude the spectra region potentially affected by absorption of the intergalactic medium. We derive the error by applying the Monte Carlo approach described in Sect. \ref{sec:integrated}. For the integrated spectrum we measure a value of -2.0 $\pm$ 0.1. This slope is not very extreme compared to other galaxies at similar redshift \citep[e.g.][]{Cullen23,Topping24,Tang25,Donnan25} and dominated by the flux of the central clump which contains already a somewhat older stellar population. The $\beta$ slope measured for the central clump is very similar to the integrated value (-2.12 $\pm 0.05$). The $\beta$ slope of the tail region however, is  redder (-1.6 $\pm$ 0.3). As this region is very bright in nebular emission, the presence of nebular continuum in the UV would impact the measured $\beta$ slope by making it significantly redder \citep{Narayanan25,Katz25}. 

A signpost of a slightly older population on the other hand is the Balmer break at 4000 \AA\ \citep[e.g.][]{Wilkins23}. Following \citet{Binggeli19} and \citet{Endsley24}, we estimate the strength of the Balmer break as B$_{4200/3500}$ =  F$_\nu$ (4200 \AA) / F$_\nu$ (3500 \AA). Following \citet{Messa25} we calculate  F$_\nu$ (3500 \AA) by integrating between 3290 and 3580 \AA\ and F$_\nu$ (4200 \AA) between 4040 and 4330 \AA\ in order to make it less sensitive to spurious pixels in the spectrum. We find B$_{4200/3500}$ = 1.1 $\pm$ 0.1 from our integrated spectrum, as well as the central clump. Comparing this value to the model predictions of \citet{Wilkins23}, a Balmer break strength of unity would be consistent with a constant star formation history of $\sim100$ Myr, while in case of a single burst a Balmer break of unity is reached in $\sim 6$Myr for a metallicity of 5\% solar, the best matching metallicity in their study. The continuum emission of the entire galaxy is dominated by the central clump emission and the spectrophotometric SED fitting of the central clump by \citet{Boyett24} reveals a value of $120$ Myr for the stellar age. This would suggest that star formation in Gz9p3 began around 400 Myr after the Big Bang.

The tail region shows evidence of a reversed Balmer break (B$_{4200/3500}$ = 0.4 $\pm$ 0.3), suggestive of a very young population ($<$ 1 Myr), confirming the findings already discussed above.

\begin{figure}[!t]
\includegraphics[width=\hsize]{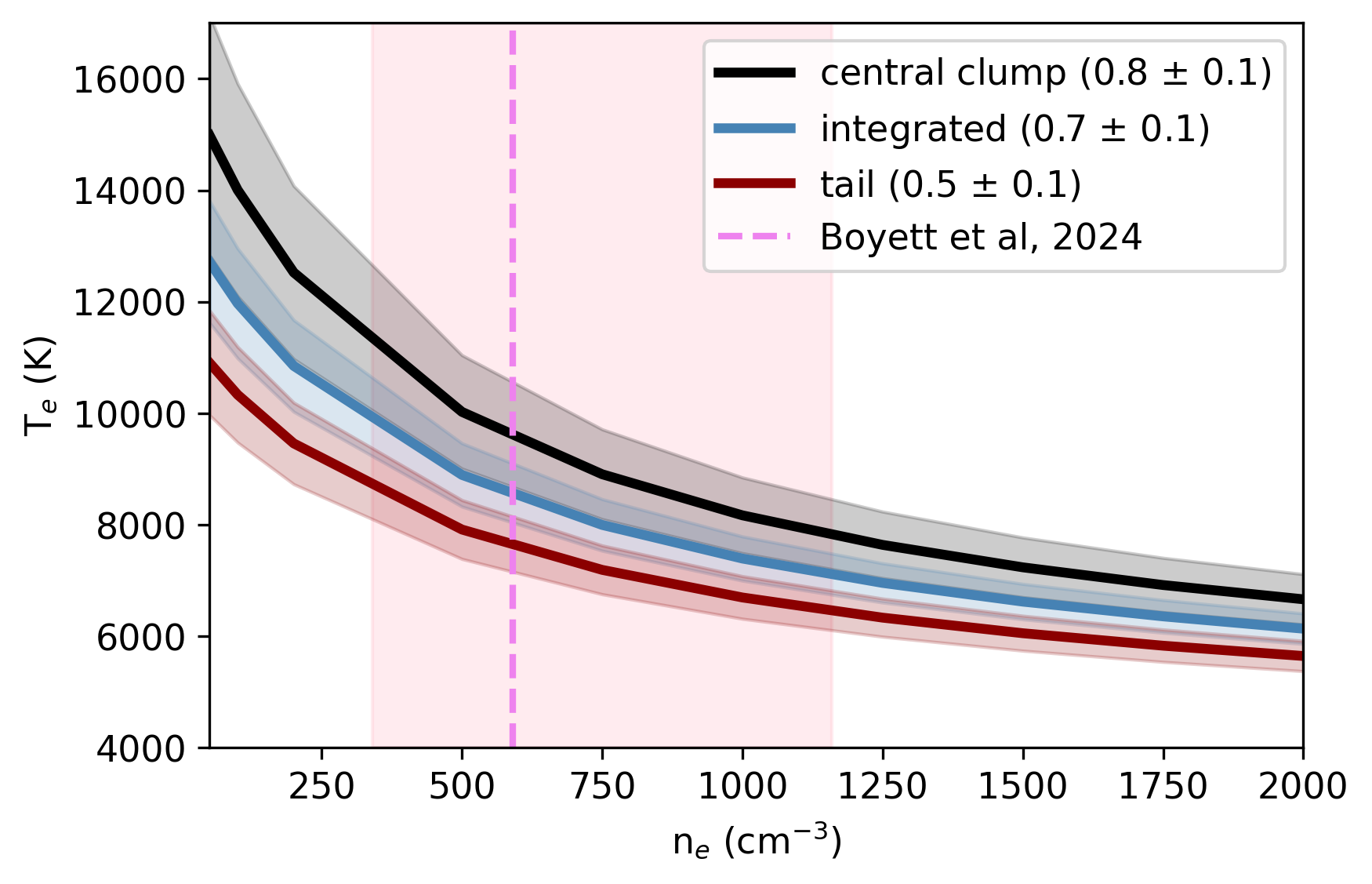}
\caption{Electron temperature and density derived with Pyneb for the observed log([OIII]$_{5008\AA}$ / [OIII]$_{88\mu m}$) ratio for the integrated aperture (blue), the central clump (black) and the tail region (red). The electron density measurement from \citet{Boyett24} is plotted as a vertical dashed line, with the error as vertical shaded area. \label{fig:OIIIratio}}
\end{figure}

\begin{figure*}
\includegraphics[width=\hsize]{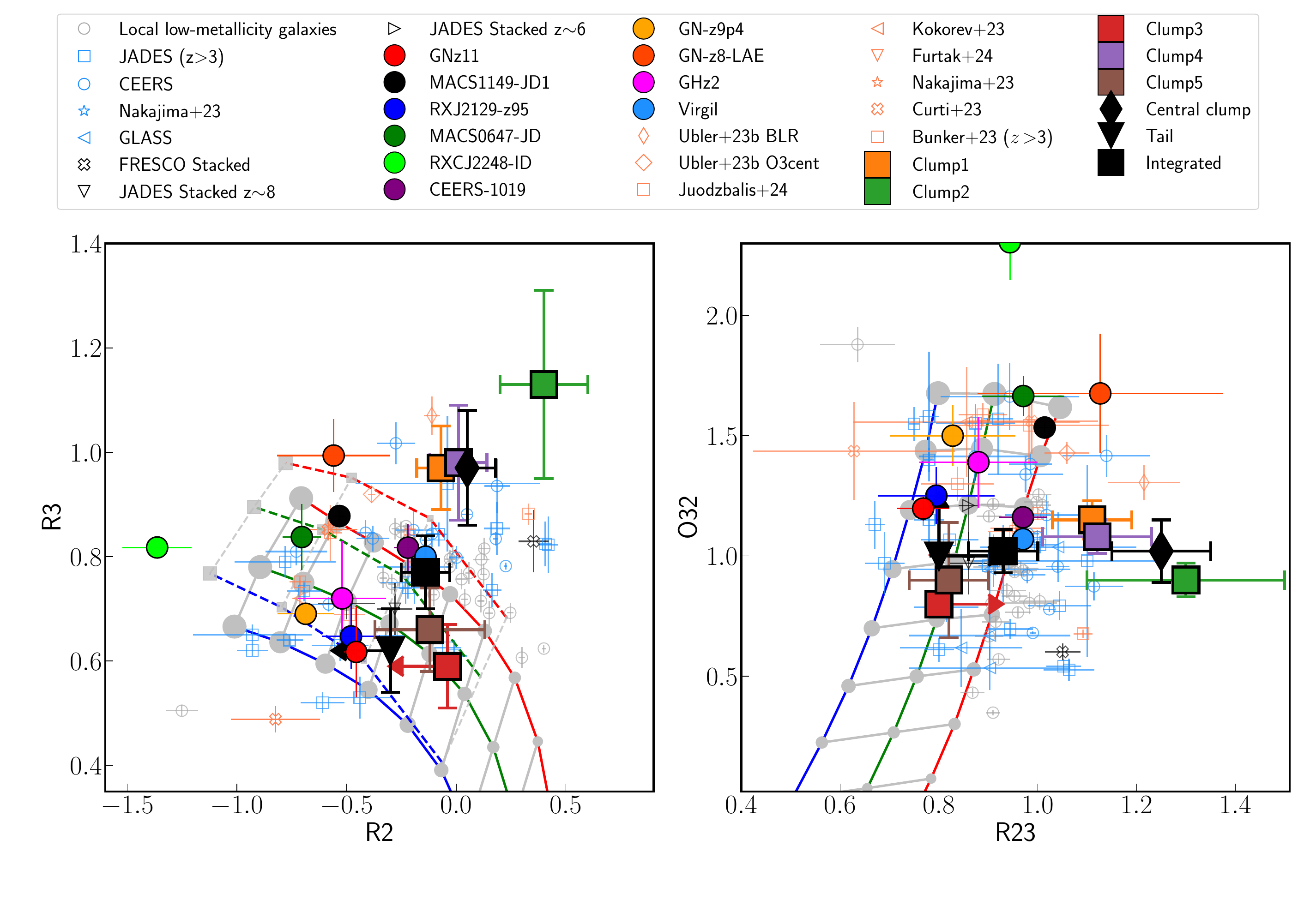}
\caption{R3 vs R2 and O32 vs R23 line ratio diagrams adapted from \citet{Alvarez25}. The integrated line ratio (Tab. \ref{tab:linefluxes}) of Gz9p3 are shown as black square, the central clump and tail region shown as black and  diamond and triangle. The spatially resolved line ratios of the 5 extracted clumps (Tab. \ref{tab:resolved_fluxes}) as colored squares. Plotted are the location of individual galaxies above $z$ = 8: GN-z8-LAE \citep{Navarro24} at z=8.73, CEERS-1019 \citep{Marques24} at z=8.7, MACS1149-JD1 \citep{Stiavelli23} at z = 9.1;  GN-z9p4 \citep{Schaerer24}  at z = 9.4, RXJ2129-z95 \citep{Williams23} at z = 9.5, MACS0647-JD \citep{Abdurrouf24} at z = 10.2, GNz11 \citep{Alvarez25} at z=10.6 and  GHz2 \citep{Calabro25} at  z = 12.3. We also included the location of Virgil \citep{Rinaldi25}, a obscured AGN hidden in a star forming galaxy at z = 6.6.
The blue markers represent high-z star forming galaxies above $z$ = 3 from JADES \citep[both individual and stacked subsamples at z$\sim$ 6 and 8,][]{Bunker23,Cameron23}, CEERS \citep{Sanders24}, GLASS \citep{Mascia23}, FRESCO \citep{Meyer24} as well as sources from \citep{Nakajima23}. The red markers represent the location of a sample of $z$ $>$ 5 type 1 AGNs \citep{Bunker23,Curti23,Kokorev23,Nakajima23,Furtak24,Juodzbalis24,Ubler24}. The black symbols are a sample of low-redshift metal-poor (12 + log(O/H) $<$ 8) star forming galaxies \citep{Izotov06,IzotovThuan11,Izotov16,Izotov24}. We overlay photoionization models from \citet{Vale-Asari16} for the metallicities 12+log(O/H) = 7.6 (blue), 7.8 (green) and 8.0 (red) in solid lines. The tracks are plotted for a starburst age of 4 Myr and a range of ionization parameters, log(U) = -1  to -4, shown by the size of the circle. The dashed lines are AGN tracks from \citet{Calabro23} for the metallicities 12+log(O/H) = 7.67 (blue), 7.84 (green), 7.97 (red).
\label{fig:R2R3}}
\end{figure*}

\subsection{Electron temperature and density}\label{sec:OIIIappendix}

The ratio of the optical rest frame [OIII] $\lambda$5008 and [OIII] line at 88.4 \micron\ is sensitive to the density and temperature of the ionized ISM \citep[e.g.][]{Stiavelli23,FujimotoALMA24,Usui25}. The spatial distribution of both [OIII] lines is remarkably different. As discussed in Sect. \ref{sec:linemaps}, the [OIII] emission at 5008\AA\ peaks towards the clump 1 aperture, south-west of the continuum bright clump. A second peak in the optical [OIII] emission is seen towards the faint UV tail. The ALMA [OIII] emission, however, in between the central clump and the tail region, suggesting changing physical conditions (temperature and density) with location in the ISM of Gz9p3.

In order to quantify these different physical conditions, we calculate the ratio for Gz9p3 by combining the [OIII]$\lambda$5008 emission line map extracted from our NIRSpec data with the [OIII]$\lambda$88\micron\ map from \citet{Algera25}.  We construct a [OIII]$\lambda$5008 emission line map following the procedure described in Sect \ref{sec:linemaps} (Note the [OIII] emission line map displayed in Fig. \ref{fig:OIIImap} is the sum of the 5008\AA\ and 4960\AA\ emission.) We convert the ALMA [OIII] emission line map to the same flux units as the NIRSpec [OIII] map. We construct a convolution kernel to match the PSF of the NIRSpec to the beam size of the ALMA map assuming both PSFs are gaussian shaped. For the NIRSpec PSF we use the formulism by \citet{Deugenio24}, 0.22\arcsec $\times$ 0.19\arcsec at the observed wavelength of [OIII]. The beam size of the ALMA data is 0.72\arcsec $\times$ 0.58\arcsec.
After convolving the NIRSpec [OIII] linemap with the kernel we extract the flux in both line maps in the three apertures used in the paper, the integrated, the central clump and the tail aperture. The measured ratios log([OIII]$_{5008\AA}$ / [OIII]$_{88\mu m}$) are given in Table. \ref{tab:linefluxes}. For the integrated aperture we derive a ratio of 0.7$\pm$0.1, while for the central clump and tail regions we derive ratios of 0.8$\pm$0.1 and 0.5$\pm$0.1 respectively.

As this line ratio is sensitive to both the electron temperature ($T_{\mathrm{e}}$) and electron density ($n_{\mathrm{e}}$), there is no unique solution for a measured line ratio. Instead we calculate the expected  $T_{\mathrm{e}}$ for a range of electron densities between $n_{\mathrm{e}}$ = 100 and 3000 cm$^{-3}$ using \texttt{getTemDen} from Pyneb \citep{Luridiana15}. The densities are chosen to reflect the typical densities observed in high redshift galaxies \citep[e.g.][]{Abdurrouf24}. 
The errors on the flux values are calculated from the standard deviation of sky apertures place around the galaxy, as described in Sect. \ref{sec:integrated}. As Pyneb does not do error propagation we calculated the error on the $T_{\mathrm{e}}$ estimates using a Monte Carlo simulation by creating 1000 realizations of the observed ratio by taking into account the error on the flux values. The resulting values and their errors are plotted as shaded bands in Fig. \ref{fig:OIIIratio}. We find a temperature range of $T_{\mathrm{e}}$ = 7000 - 16000 K, where the higher values for $T_{\mathrm{e}}$ are for the lower densities (Fig. \ref{fig:OIIIratio}). The derived temperatures are in the range of [OIII] temperatures derived using from the [OIII]4364\AA\ auroral emission line in other galaxies \citep[e.g.][]{Sanders24}. These results show densities and temperatures typical for star forming galaxies at high redshift, where lower metallicity galaxies should have slightly higher $T_{\mathrm{e}}$ and lower $n_{\mathrm{e}}$ \citep{Nicholls00}.

The vertical line in Fig. \ref{fig:OIIIratio} shows the measured electron density  $n_{\mathrm{e}}$ = 590$^{+570}_{-250}$ cm$^{-3}$ based on the flux ratio of the resolved [OII] doublet \citep{Boyett24}. The MSA slit from which this spectrum was extracted was centered on the central clump of the galaxy allowing us to constrain the $T_{\mathrm{e}}$ for the central clump slightly better. Adopting this value for $n_{\mathrm{e}}$ for the central clump would result in a estimate for $T_{\mathrm{e}}$ =  9500$^{+3500}_{-2000}$ K. The observed ratio for the tail region is lower. Considering its lower metallicity, this could indicate a lower $n_{\mathrm{e}}$ and, possible, higher temperature relative to the central clump of the galaxy.

\subsection{Emission line ratio diagrams}\label{sec:emissionratios}

In Figure \ref{fig:R2R3} we present the R3-R2 and O32-R23 emission line diagrams which  trace the ionization and excitation conditions of the ionized gas \citep[e.g.][]{Kewley19}. We compare the location of the integrated line ratio of Gz9p3 (black square) and the different apertures to other high redshift star forming galaxies and AGNs, local low-metallicity extreme starforming galaxies as well as photo-ionization and AGN models. The location of Gz9p3 shows that as for many high-redshift galaxies, the gas is highly ionized, showing a high O32 ($\sim1$) and R3 ($\sim0.8$) ratio  \citep[c.f.][]{Langeroodi23,Cameron23}. From the O32 value of Gz9p3 we calculate the ionization parameter $U$ using the calibration of \citet{Berg19} and find log(U) = -2.0 $\pm$ 0.1 for the integrated spectrum, as well as the central aperture. For the tail region we derive a lower limit on the O32 index due to the faint [OII] emission, resulting in a lower-limit of log(U) = -2.0. These values show that the ionization in the galaxy is rather homogeneous, and likely slightly higher in the tail region, hosting the youngest stellar population.

We also show the location of the central clump and the tail region (black diamond and triangle) as well as the five extracted NIRSpec only clumps plotted with colored squares in both diagrams. The points show a large range on observed R3 and R23 indices. The tail of the galaxy  have R3 ratios below 0.7 (clumps 3 and 5), showing bright \hb\ in their spectrum (Fig. \ref{fig:clumpspectra}). The central clump and other three clumps (1,2 and 4) on the other hand have R3 ratios close to unity and an R23 ratio above unity. Especially clump 2 is an outlier in the line ratio plots, although with large errors, occupying parameter space where not many galaxies are found.  This would mean that the excitation of the gas as traced mostly by the R23 index \citep{Kewley19} differs quite strongly between the different clumps.

We compare the location of these points with other high redshift star forming galaxies (blue symbols) and type 1 AGNs (orange symbols) in the two line ratio diagrams (Fig. \ref{fig:R2R3}). There is substantial scatter and overlap between the star forming galaxies and the type 1 AGNs in these plots. However, the type 1 AGN typically occupy the upper regions of the R3-R2 plot (higher R3) and the top-right part of the O32 - R23 diagram (high R23 and high O32).

%Finally, we show the location of photo ionization model tracks for metallicities similar to Gz9p3 from \citet{Vale-Asari16} in both pannels as solid lines. The location of these tracks corroborates the observation that star forming galaxies are typically observed in the bottom left of the R2-R3 and at R23 values below unity. The models also show that the O32 index strongly correlates with the ionization parameter U as shown by the size of the plum circles. 

%We plot the AGN model tracks for metallicities similar to Gz9-3 from \citet{Calabro23} as dashed lines in the R3 vs R2 diagram. It becomes clear that there is very little difference between the AGN and star formation tracks. \citet{Cleri25} demonstrate that at low metallicity and high log(U), there is very little difference in line ratios between AGN and star formation models.

\section{Discussion}\label{sec:discussion}

\subsection{Does Gz9p3 harbor an AGN?}

In the spatially resolved analysis of Gz9p3 we find high R3 and R23 values for the central clump that are offset compared to other star forming galaxies and more consistent with the location of AGN. In order to identify the mechanism driving the high R3 and R23 indices, we compare the location of Gz9p3 in the diagrams in Fig. \ref{fig:R2R3} with photo ionization model tracks for starbursts from \citet[solid lines][]{Vale-Asari16} and AGN model tracks from \citet[dashed lines][]{Calabro23}. 

Comparing the location of the two model tracks show  that there is very little difference between the AGN and star formation tracks at low metallicity  (12+log(O/H) = 7.6 - 8.0).  \citet{Cleri25} demonstrate that at low metallicity and high log(U), the line ratios between AGN and star formation models occupy the same space in the line ratio diagrams such as in Fig. \ref{fig:R2R3}.  Higher  metallicity AGN tracks do go to higher R3 ratios \citep[e.g.][]{Feltre16,Calabro23}, explaining also measurements of the AGN in the literature overplotted in Fig. \ref{fig:R2R3}. 

Under the adoption of the measured metallicity, it becomes clear that of the derived line ratios in especially the central clump are difficult  to reconcile with either a star formation  or AGN driven excitation of the ISM of Gz9p3. We explored the possibility of shock excitation playing a role in explaining the observed line flux ratios. Shocks have been observed in nearby merging galaxies \citep[e.g.][]{Rich15} based on spatially resolved IFU observations. The shock+precursor models from \citet{Allen08} and \citet{Flury25} for the models with metallically of that of Gz9p3 do not show these extreme values R3 values as observed either. 

\begin{figure}
\includegraphics[width=\hsize]{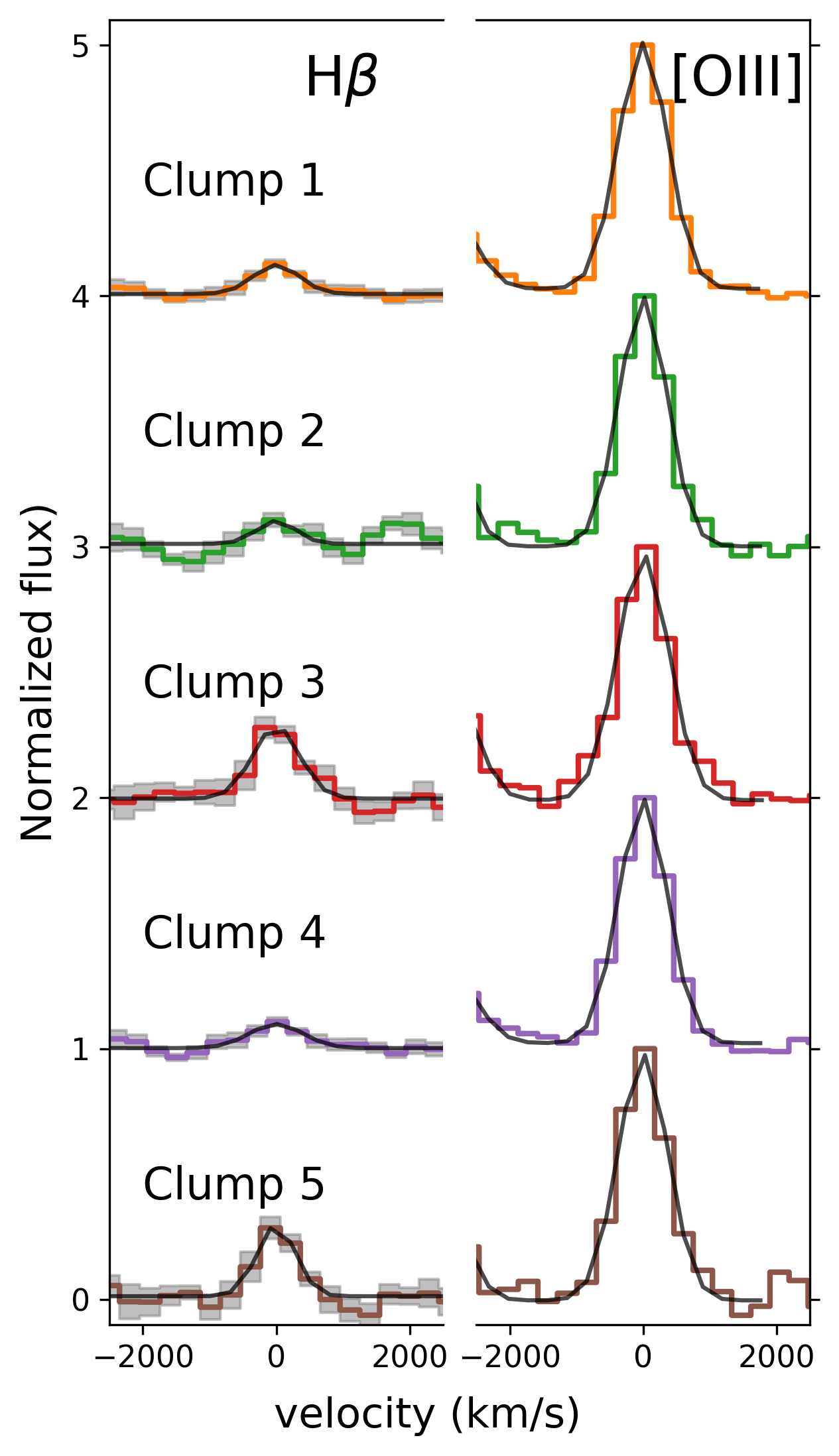}
\caption{Zoom in of the NIRSpec spectrum showing \hb\ (left) and [OIII]5008\AA\ (right) emission lines for each of the clumps. Overplotted are the best gaussian fits to the emission lines. No evidence for broad \hb\ is present in the spectra.
\label{fig:Hbetaspec}}
\end{figure}

We searched for the presence of [NeV] at a rest wavelength of 3426\AA. This line could be another signpost for shocks \citep{Allen08} or  AGN \citep{CleriNeratio23,Cleri23}, where the strength of the [NeV] line relative to [NeIII] can be used as a discriminator. At first glance we detect a spectral feature at the wavelength of the [NeV] emission line in both the spatially integrated emission as well as the spectrum of the central clump. Inspection of the spatial location revealed that the [NeV] emission was not centered on the continuum emission of the galaxy, After a more detailed analysis and reprocessing of the data using a different reduction software \citep{Perna23}, we conclude that the [NeV] is contaminated by an artifact and not significant, placing an upper limit on the detection as given in Table \ref{tab:linefluxes} and \ref{tab:resolved_fluxes}. % In appendix \ref{sec:NeV} we describe the procedure in detail.% The upper limits given to the [NeV] fluxes in Tables \ref{tab:linefluxes} and \ref{tab:resolved_fluxes} should be interpreted as conservative upper limits.

%In the nearby universe, [NeV] is detected in blue compact galaxies, low redshift star forming galaxies \citep{Thuan05,Izotov12}.  After comparing the observed line ratios with shock models from \citet{Allen08}, these studies conclude that the [NeV] is originating from fast shocks.
%However, the strength of the [NeV] line relative to \hb\ is several orders of magnitude fainter than in Gz9p3. \citet{Cleri23} analyze a sample of [NeV] emitters between redshift 1.4 and 2.3 and concluded based on line ratio diagrams that these [NeV] emitters are most likely AGNs. 

\citet{CleriNeratio23} construct a line ratio diagram using the O3 index and the log([NeV]/[NeIII]) line ratio (Ne53) to descriminate between galaxies dominated by AGN or star formation as well as a composite of AGN and star formation. The R3 index and the NeV upper limits of Gz9p3 are given in Tab. \ref{tab:linefluxes}. We calculate a Ne53 index for the integrated spectrum (Ne53 <  0.54) and the spectrum of the central clump (Ne53 < 0.63). Combined with the high R3 index this upper limit would be consistent with an AGN driven emission, but cannot be discriminated from star formation as we do not have a significant detection of [NeV].

Many type 1 AGN at high redshift are identified based on broad wings under the \hb\ and \ha\ recombination lines with velocities of several  1000 \kms \citep[e.g.][]{Furtak24,Ubler24,Juodzbalis24}, while the forbidden emission lines such as [OIII] remain narrow. The MIRI/MRS spectrum (Fig \ref{fig:MRS_integrated}) provides the highest spectral resolution spectrum available for Gz9p3. The \ha\ spectrum does not show evidence for a broad underlying component, but the SNR of the spectrum is too low to make a strong statement about it.
The \ha\ profile shows some velocity structure at low significance in the form of a blue wing, this wing originates from the clumps in the tail of the galaxy.

In Fig. \ref{fig:Hbetaspec} we show the NIRSpec spectra of \hb\ and [OIII] from the extracted clumps. Overplotted are the best fits to the emission lines as described in Sect. \ref{sec:integrated}. We do not find any difference in the fitted linewidth of \hb\ between the different clumps. We also do not find any significant difference between the \hb\ line width and the [OIII] line width and therefore no signature for AGNs based on the line with. Note that the spectral resolution of our  NIRSpec spectra is low and the \hb\ lines in clump 1, 2 and 4 are not very strong and a faint broad component would be easily missed in the spectrum. High spectral resolution and deeper spectra would be needed to provide a more conclusive answer. 

Similar results were found by  \citet{Fujimoto24} with MSA-slit spectra, who measured an R3 ratio above unity. The orientation of the MSA slit was along the galaxy, with most of the emission coming from clumps 1, 2 and 4. Based on the non-detection of the MgII line (also not detected in our spectra), as well as the rest-frame UV-optical morphology from the NIRCam images and the SED fitting, \citet{Boyett24} on the other hand concluded that there is no evidence for the presence of an AGN.

Combining our observations with the previous studies would suggest that a hard source of radiation is present, providing enough high energy photons to alter the observed line ratios without dominating the rest frame UV and optical morphology. Merging galaxies are known to result in an increase of star formation and the creation of black holes in the center.

Observations at longer wavelengths with e.g. MIRI would help to shed light on what causes the extreme line ratios. A dusty type 2 AGN embedded in a star forming galaxy will not be visible at UV and optical rest frames, but only become visible at longer wavelengths. In Fig. \ref{fig:R2R3} we overplot the location of Virgil, a normal star forming galaxy in the optical where long wavelength MIRI imaging reveals the presence of a type 2 AGN \citep{Iani25,Rinaldi25}.
Additionally, recently \citet{CrespoGomez26} observed an infrared excess for GNz11, hinting at the presence of an obscured AGN.

Our understanding of how black holes form in the early universe and how they grow is far from complete \citep[e.g.][]{Inayoshi20}. JWST is providing more and more observational evidence that AGN are present already very early in the universe.
Gz9p3 is one of the few galaxies at z > 9 where evidence for the presence of an AGN is presented. \citet{Taylor25} identify a little red dot (LRD) at redshift 9.288 revealing a broad \hb\ line profile and narrow [OIII] emission lines, indicative of emission from the broad line region of an AGN. 
\citet{Curti25} report a detection  of [NeV] at 3426\AA\ at 3$\sigma$ significance in a $z$ = 9.4 star forming galaxy, while the other emission lines in the galaxy are consistent of star formation, suggestive of a type 2 AGN. 

At higher redshifts, the presence of a type 1 AGN has been claimed in GNz11 based on the presence of some UV lines and broad components in Balmer lines \citep{Maiolino24}. However, the optical and red spectrum of this galaxy appears to be dominated by a low-metallicity starburst \citep{Alvarez25} while the emission from a type 2 dusty torus could be present in the red and near-IR range \citep{CrespoGomez26}.

%I think this 2D burstiness difference is very interesting as shows dramatic differences in the burstiness within the same system at these redshifts indicating that while some regions are evolving in a more quiescent, long term phase, others, like in the tail, are showing a rather extremely bursty phase, may be as a resulst of the merging process (similar Halpha enhanced clumps in the tails of low-z mergers have been detected in ULIRG systems: Monreal-Ibero+2007, A&A, 472, 421. I think we could discuss part of this discussion in section 5.2

%I would consider using the term "lull-phase" if we mean that such clumps are below what it is expected for a MS galaxy; quiescent calls in a scenario where the SF has been bascially suppressed; see Looser+

%or another scenario: gas in the outskirts of the galaxies moving inwards and  still not reprocessed?
%; metallcitiy can get dilluted here
%this is also very interesting. At face value, large differences in metallicity exist within a system and on scales on kpc, Also, it seems that the regions showing higher metallicity are the ones with lower burstiness. So, the forward scenario is the presence of very bursty recent SF in a poor-metal ISM environment (tip of the tail) while a more constant and older  SF appears in a more metal enriched ISM is present in the central regions (coalescence of the two galaxies). So, a discussion along these lines could be included in section 5.2?

\subsection{Star formation in Gz9p3}
In this paper we have derived several star formation tracers for Gz9p3, both integrated over the entire galaxy and spatially resolved, allowing us to constrain the star formation properties in Gz9p3 and compare it to the properties derived from (mostly) imaging studies of this galay.  Our spatially resolved NIRSpec and MIRI/MRS spectra provide for the first time a spatially resolved view of the conditions in the ISM of this galaxy. 

The high observed integrated EW([OIII]) makes Gz9p3 an extreme emission line galaxy \citep[EELG, e.g.][]{Boyett24EELG}. These are galaxies whose emission is dominated by a young intense starburst, reflecting also the high ionization of this galaxy as traced by the high O32 ratio. On the other hand, the observed EW(H$\beta$) 
and EW(H$\alpha$) of Gz9p3 (Tab. \ref{tab:linefluxes}) are not extremely high compared to other (high redshift) galaxies \citep[e.g.][]{Alvarez24,PrietoJimenez25}. Additionally, the ionizing photon efficiency of Gz9p3 is in line with other high redshift galaxies. 

The values derived for the spatially resolved apertures show a large diversity. The strongest nebular emission comes from clump 1, the small clump just south-west of the main central clump. The star formation derived from the H$\alpha$ line suggests a rather equal distribution between the central clump aperture (6.1 $\pm$ 1.3 M$_{\odot}$ yr$^{-1}$)  and the tail aperture (6.5 $\pm$ 1.3 M$_{\odot}$ yr$^{-1}$). The distribution of the rest frame UV and optical continuum however, is very different.  The central clump of the galaxy is the brightest source in the UV and optical continuum, while the tail region is barely visible in the JWST images. \citet{Boyett24} measure an M$_{UV}$ = -18.2, compared to M$_{UV}$ = -21.1 for the entire galaxy, showing that only 7\% of the UV luminosity is emitted in the tail region. The F150W image (panel b, Fig. \ref{fig:OIIImap}), tracing the 1500\AA\ rest frame, shows only a few small clumps in the tail region, while there is no continuum source detected in the center of the [OIII] line emission towards the tail.

The photometric analysis of \citet{Boyett24} shows that the central region of the galaxy hosts a young stellar population ($<$ 50 Myrs), surrounded by an older population (50 - 150 Myrs). The clump detected in the tail region also shows evidence for a young ($<$ 50 Myrs) stellar population. 

Our spectroscopic measurements emphasize the difference in age between the tail and the clump even stronger. 
The tail region  shows a very high $\xi_{ion}$ (26.1 $\pm$ 0.2), suggestive for a very young and extreme stellar population, while the central clump shows a $\xi_{ion}$  below the canonical value (25.0 $\pm$ 0.1). The tail region has a very high burstiness parameter (10 times higher than the central clump), indicating that star formation only started recently in this region of Gz9p3. Regions with even lower UV luminosities than the tail region in Gz9p3 have been found by studying strongly lensed galaxies \citep[e.g.][]{Vanzella24,MessafaintUV25}, possibly forming massive star clusters as detected in the early universe \citep[e.g.][]{Adamo25}. On the other hand, the  burstiness parameter of the central clump is below unity (0.5 $\pm 0.2$).  The central clump in Gz9p3 is no longer the main site of current star formation, most star formation has seized with the majority of the stellar population formed over the last 50-150 Myrs. 

Additionally, we find a metallicity gradient in the galaxy where the tail region has a 0.5 dex lower metallicity compared to the central clump. This could mean that the tail region is consisting of more pristine gas, while the gas in the central clump has been chemically enriched by older generation of stars.

These results demonstrate two very different star formation environments in Gz9p3;  the presence of very bursty recent SF in a metal-poor ISM environment (tail region) while a more constant and older  star formation appears in a more metal enriched ISM is present in the central clump. Under the assumption that this galaxy is a merging galaxy, we could speculatively relate the age of the stellar population in the central region to the time of the merger event, triggering a burst of star formation with a current age of 50 - 150 Myrs, suggesting a merger event around z$\approx$12.

\subsection{Gz9p3 in the context of mergers at low- and high-z}

Based on the morphological structure, Gz9p3 has been identified as a system in a close interaction/merging phase \citep{Boyett24}. Note, however, kinematic evidence to support the merger morphology is not available and galaxies at high redshift in general appear clumpy and irregular.
Systems involving multiple mergers \citep[A2744][]{Hashimoto23}, massive dusty star-forming interacting galaxies in overdensities \citep[SPT0311-58][]{Arribas23,Alvarez23,Spilker22}, and advanced interacting galaxies \citep[B14-65666][]{Sugahara25,PrietoJimenez25,Jones26,Hashimoto23B14}
 have been identified at somewhat lower redshifts, during the Epoch of Reonization (z$\sim$6-8, i.e. 0.65-1 Gyr after the Big Bang). Gz9p3 is a member of this family of high-z interacting systems demonstrating the presence and role of gravitational interactions during the early phases of galaxy formation, even before the main EoR, when the Universe was just about 0.5 Gyr. 

The overall properties of Gz9p3 (SFR, $\xi_{ion}$, extinction, excitation conditions, metallicity, etc)  are similar to the general population of galaxies in the EoR. However, our new integral field spectroscopy combined with existing imaging has shown a complex 2D structure on (sub)kpc-scales. The stellar and ionized ISM structure consists of two well identified regions, the main body of the system and the long tidal tail. The main body consists of two bright UV-continuum clumps that could be considered as the nuclei of the two proto-galaxies involved in the interaction. The UV-faintest of the two is dominating both the [OIII] and H$\alpha$ emission. This indicates that while the two proto-galaxies are in the advanced coalescence phase, one is in a more active phase (either star-forming or AGN), while the
other is in a more quiescent phase, likely associated with an older (100-120 Myr) stellar population. The tail region of the ionized ISM show on the other hand an [OIII] and H$\alpha$ emission characterized by very high equivalent widths typical of very young starbursts with almost no previous star formation. So, this could be interpreted as an on-going in-situ star formation along the tail. These features are common in low-z interacting systems where star formation traced by the H$\alpha$ are detected along the tidal tails \citep[e.g.][]{Arribas00, Monreal07,Knierman13}, forming bound star clusters \citep[e.g.][]{Rodruck23}.

While the star-formation in the main body has a low metallicity of about 0.18 times the solar metallicity, the star formation in tail is proceeding in a  metal-poor environment 0.06 solar (see Sect. \ref{sec:metallicity}). This has important implications as it would allow to investigate the how the star formation proceeds at very early epochs in gas rich mergers with very different metallicity environments within the same system, as in Gz9p3.

Recent simulations of the formation of galaxies during the EoR  show that systems identified as clumpy – i.e., with two or more high-surface-density star-forming clumps as traced by emission lines – represent 10\% of all simulated sources \citep[and references therein]{Nakazato24}. Most of these clumpy
systems are produced by mergers, and represent a short-lived, transient phase with the young star formation along the gas-rich tidal tails, and with specific SFR of the order of 50 M$_{\odot} $Gyr$^{-1}$, before final coalescence into a single system. Thus, the study of systems like Gz9p3 provides  unique opportunities to investigate the formation and evolution of galaxies, and their stellar mass assembly, as a consequence of mergers in the early universe. Additional deep JWST integral field spectroscopy  of other systems like Gz9p3 are required for this purpose.

\section{Conclusions}\label{sec:conclusions}

In this paper we combine deep NIRSpec-PRISM and MIRI/MRS integral field spectroscopy to characterize the spatially resolved ISM in the EoR galaxy Gz9p3 at $z$ = 9.3127. We combine this analysis with MIRI F560W and F770W imaging and archival NIRCam imaging \citep{Boyett24}. This is only the third galaxy at z>9 where such a dataset is available. We detected spatially resolved nebular lines and we extract spectra using different apertures to characterize the spatially resolved ISM properties. Our findings can be summarized as follows:
\begin{itemize}
    \item{We detect spatially resolved \ha\ emission with MIRI-MRS at z=9.3127 $\pm$ 0.0002.}
    \item{The spatial distribution of the ionized gas, traced by the [OIII] emission shows a clumpy nature, like the continuum imaging. The brightest [OIII] emission originates from the central clump area. However, a second peak of [OIII] emission is detected towards the UV faint clump region.}
    \item{We measure the Balmer decrement and find it consistent with no extinction. This is consistent with the values found based on the 2D SED fitting of the NIRCAM images.}
    \item{We detect large spatial variations of $\xi_{ion}$, burstiness of star formation and metallicity between the UV bright central clump and the tail region indicative of a large diversity of ISM and star formation properties in the same galaxy.}
    \item{We measure spatial differences in the ratio of [OIII] 5008\AA\ line with respect to the [OIII] 88\micron\ lines, which we interpret as indication different physical conditions in the ISM (lower density) in the tail region compared to the central clump.}
    \item{We identify the tail region as a very recent low-metallicity starburst and possible formation site of massive star clusters. The central clump instead is forming stars more constantly and has formed most of its stars 50 -150 Myrs ago and has a more enriched ISM.}
    \item{We find evidence for extreme excitation conditions in the central clump, However, star formation, kow-metallicity AGN or shock models cannot explain the observed line ratios. We searched for other evidence supporting non-stellar excitation but found no conclusive detection of [NeV] or broad Balmer lines.}

\end{itemize}

This study demonstrates the power of spatially resolved spectroscopy with JWST allowing us to study the individual clumps and the spatial variations in the ISM in even the highest redshift galaxies. This is particular true when dealing with dynamically young interacting/merging proto-galaxy systems in the very early Universe.

\begin{acknowledgements}
The authors thank Almudena Alonso Herrero for reading an earlier version of the manuscript. A.B., G.O. and J. M. acknowledge support from the Swedish National Space Administration (SNSA). J.A.-M. acknowledges support by grants PID2024-158856NA-I00 \& PIB2021-127718NB-I00 from the Spanish Ministry of Science and Innovation/State Agency of Research MCIN/AEI/10.13039/501100011033 and by “ERDF A way of making Europe”. L.C. acknowledges support by grant PIB2021-127718NB-100 from the Spanish Ministry of Science and Innovation/State Agency of Research MCIN/AEI/10.13039/501100011033 and by “ERDF A way of making Europe”. HSBA gratefully acknowledges support from Academia Sinica through grant AS-PD-1141-M01-2.
L.A.B. acknowledges support from the Dutch Research Council (NWO) under grant VI.Veni.242.055 (\url{https://doi.org/10.61686/LAJVP77714}).This work was supported by research grants (VIL16599,VIL54489) from VILLUM FONDEN. 
TRG acknowledges funding from the Cosmic Dawn Center (DAWN), funded by the Danish National Research Foundation (DNRF) under grant DNRF140. TRG is also grateful for support from the Carlsberg Foundation via grant No. CF20-0534.
M.P. acknowledges support through the grants PID2021-127718NB-I00, PID2024-159902NA-I00, and RYC2023-044853-I, funded by the Spain Ministry of Science and Innovation/State Agency of Research MCIN/AEI/10.13039/501100011033 and El Fondo Social Europeo Plus FSE+. J.P. acknowledges financial support from the UK Science and Technology Facilities Council, and the UK Space Agency.

The work presented is the effort of the entire MIRI team and the enthusiasm within the MIRI partnership is a significant factor in its success. MIRI draws on the scientific and technical expertise of the following organisations: Ames Research Center, USA; Airbus Defence and Space, UK; CEA-Irfu, Saclay, France; Centre Spatial de Liége, Belgium; Consejo Superior de Investigaciones Científicas, Spain; Carl Zeiss Optronics, Germany; Chalmers University of Technology, Sweden; Cosmic Dawn Center (DAWN), DTU Space, Technical University of Denmark, Denmark; Dublin Institute for Advanced Studies, Ireland; European Space Agency, Netherlands; ETCA, Belgium; ETH Zurich, Switzerland; Goddard Space Flight Center, USA; Institute d'Astrophysique Spatiale, France; Instituto Nacional de Técnica Aeroespacial, Spain; Institute for Astronomy, Edinburgh, UK; Jet Propulsion Laboratory, USA; Laboratoire d'Astrophysique de Marseille (LAM), France; 
Leiden University, Netherlands; Lockheed Advanced Technology Center (USA); NOVA Opt-IR group at Dwingeloo, Netherlands; Northrop Grumman, USA; Max-Planck Institut für Astronomie (MPIA), Heidelberg, Germany; Laboratoire d’Etudes Spatiales et d'Instrumentation en Astrophysique (LESIA), France; Paul Scherrer Institut, Switzerland; Raytheon Vision Systems, USA; RUAG Aerospace, Switzerland; Rutherford Appleton Laboratory (RAL Space), UK; Space Telescope Science Institute, USA; Toegepast- Natuurwetenschappelijk Onderzoek (TNO-TPD), Netherlands; UK Astronomy Technology Centre, UK; University College London, UK; University of Amsterdam, Netherlands; University of Arizona, USA; University of Cardiff, UK; University of Cologne, Germany; University of Ghent; University of Groningen, Netherlands; University of Leicester, UK; University of Leuven, Belgium; University of Stockholm, Sweden; Utah State University, USA.
A portion of this work was carried out at the Jet Propulsion Laboratory, California Institute of Technology, under a contract with the National Aeronautics and Space Administration.
We would like to thank the following National and International Funding Agencies for their support of the MIRI development: NASA; ESA; Belgian Science Policy Office; Centre Nationale D'Etudes Spatiales (CNES); Danish National Space Centre; Deutsches Zentrum fur Luft-und Raumfahrt (DLR); Enterprise Ireland; Ministerio De Econom\'ia y Competitividad; Netherlands Research School for Astronomy (NOVA); Netherlands Organisation for Scientific Research (NWO); Science and Technology Facilities Council; Swiss Space Office; Swedish National Space Board; UK Space Agency. 

This work is based on observations made with the NASA/ESA/CSA James Webb Space Telescope. Some data were obtained from the Mikulski Archive for Space Telescopes at the Space Telescope Science Institute, which is operated by the Association of Universities for Research in Astronomy, Inc., under NASA contract NAS 5-03127 for \textit{JWST} Some of the data products presented herein were retrieved from the Dawn JWST Archive (DJA). DJA is an initiative of the Cosmic Dawn Center (DAWN), which is funded by the Danish National Research Foundation under grant DNRF140.

This work made use of astropy \citep{2013A&A...558A..33A,2018AJ....156..123A}, photutils \citep{Bradley24} and  specutils \citep{specutils}. For the purpose of open access, the authors have applied a Creative Commons Attribution (CC BY) licence to the Author Accepted Manuscript version arising from this submission.

\end{acknowledgements}

%\vspace{5mm}
%\facility{JWST (NIRSpec, MIRI and NIRCam)}




\bibliography{sample631}{}
\bibliographystyle{aa.bst}
\onecolumn

\begin{appendix}

%\section{Validation of potential [NeV] emission.}\label{sec:NeV}
%{\bf not yet finished}

%The spectrum of Gz9p3, presented in Fig. \ref{fig:NIRSpec_integrated}, shows evidence for the presence of [NeV] emission at a rest wavelength of 3426 \AA. Further inspection of the [NeV] emission in the datacube revealed that the emission was not spatially coinciding with the continuum emission or the [OIII] emission of the galaxy (Fig. \ref{fig:OIIImap}). Additionally, the line profile of [NeV] in the spectrum of Clump 4 (Fig. \ref{fig:clumpspectra}) shows a double peaked profile, raising the suspicion that a artifact affects the spectrum at the wavelength of [NeV].

%We ran stage 3 of the JWST pipeline 6 times with each time removing one exposure. If an artifact in one of the frames affects the [NeV] emission, this should disappear when that frame is excluded from the reduction. We then analyzed the datacubes in the  same way as the full data cube (Sect. \ref{sec:nirspecalib}), the background was subtracted and spectra were extracted in the same apertures. The only difference is that we did not match the psf to that of the wavelength of [OIII] in order to not smooth out a possible artifact.

\section{Spatially resolved emission line properties}

\begin{figure*}[!h]
\includegraphics[width=\hsize]{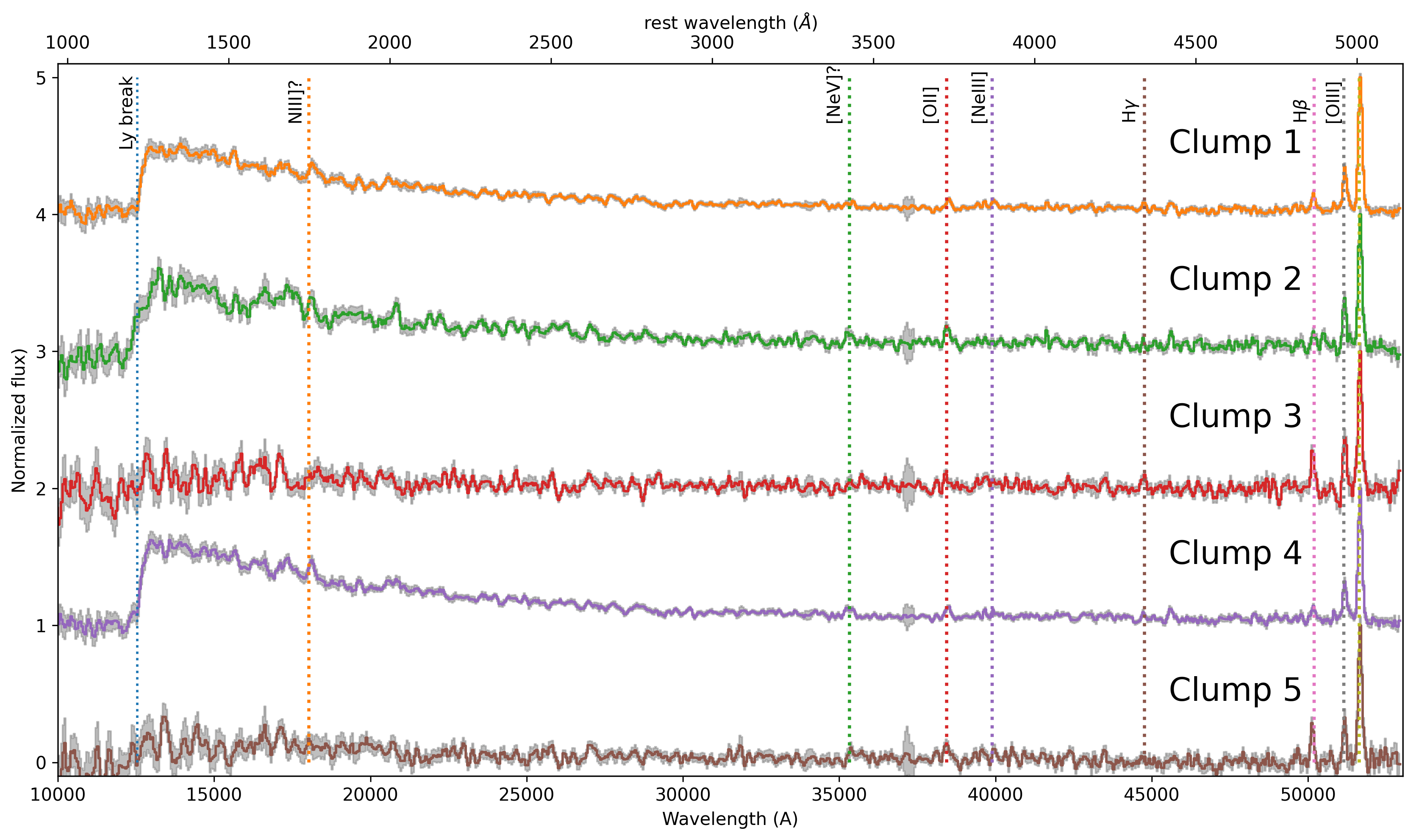}
d\caption{NIRSpec spectra of extracted in the 5 different apertures defined in Fig. \ref{fig:apertures}, normalized to the strength of the [OIII]5008 line. Clump 1-3 cover the [OIII] selected apertures and Clump 4 and 5 the continuum selected apertures.\label{fig:clumpspectra}}
\end{figure*}

\begin{table*}[!h]
\caption{Spatially resolved line fluxes and line ratios}
\centering
\begin{tabular}{rrrrrr}
Line & Clump 1 & Clump 2& Clump 3& Clump 4& Clump 5 \\
\hline\hline
%\colhead{}& \colhead{$\times10^{-19}$} & \colhead{$\times10^{-19}$}& \colhead{$\times10^{-19}$}& \colhead{$\times10^{-19}$}& \colhead{$\times10^{-19}$} 
{[OII]} $\lambda3727,3729$ & 4.6 $\pm$ 0.9 & 4.7 $\pm$ 0.7  & $<$ 5.9       & 4.8 $\pm$ 0.8 & 3.7 $\pm$ 0.2\\
{[NeIII]} $\lambda3867$  &  2.2 $\pm$ 0.6  & $<$1.2         & 2.0 $\pm$ 1.3 & 0.9 $\pm$ 0.7 & $<$0.9\\
{[NeV]} $\lambda3426$  &   $<$3.6 $\pm$ 1.9   & $<$4.0 $\pm$ 1.3  & $<$ 0.9       & $<$6.3 $\pm$ 1.1 & $<$2.0\\
H$\gamma$ $\lambda4340$ &  2.0 $\pm$ 0.6   & 0.8 $\pm$ 0.6  & 2.6 $\pm$ 0.7 & 1.4 $\pm$ 0.5 & 1.2 $\pm$ 0.9\\
H$\beta$  $\lambda4861$ & 5.4 $\pm$ 1.0    & 2.1 $\pm$ 0.8  & 6.4 $\pm$ 1.1 & 4.8 $\pm$ 1.3 & 5.0 $\pm$ 0.8\\
{[OIII]} $\lambda4960$    & 14.8 $\pm$ 1.5 & 9.4 $\pm$ 1.7  & 9.4$\pm$ 1.6  & 12.9$\pm$ 2.2 & 6.8 $\pm$ 1.8\\
{[OIII]} $\lambda5008$    & 50.7 $\pm$ 1.7 & 28.8 $\pm$ 1.7 & 25.1 $\pm$ 2.0& 45.2$\pm$ 2.0 & 23.2 $\pm$ 1.7\\
\hline
\hline
R2 & -0.07  $\pm$ 0.11   & 0.4 $\pm$ 0.2    & $<$ -0.04 & 0.01 $\pm$ 0.13 & -0.12 $\pm$ 0.25\\
O32 & 1.15  $\pm$ 0.08   & 0.90 $\pm$ 0.07  & $>$ 0.8 & 1.08 $\pm$ 0.07 & 0.90 $\pm$ 0.24\\
R23 & 1.11  $\pm$ 0.08   & 1.3 $\pm$ 0.2    & $<$ 0.8 & 1.12 $\pm$ 0.11 & 0.82 $\pm$ 0.08\\
Ne3O2 & -0.31 $\pm$ 0.15 &  $<$ -0.6            & --- & -0.7 $\pm$ 0.4  & $<$-0.6 \\
R3 & 0.97  $\pm$ 0.08    & 1.13 $\pm$ 0.18  & 0.59 $\pm$ 0.08 & 0.98 $\pm$ 0.11 & 0.66 $\pm$ 0.08\\
$\hat{R}$ & 0.82 $\pm$ 0.09 & 1.17 $\pm$ 0.19 & 0.3 $\pm$ 0.6 & 0.63 $\pm$ 0.12 & 0.53 $\pm$ 0.13 \\
\hline
\hline
$\beta_{UV}$ & -2.22 $\pm$ 0.04 & -1.87 $\pm$ 0.06 & -2.2 $\pm$ 0.3 & -2.11 $\pm$ 0.04 & -1.9 $\pm$ 0.2 \\
F(4200)/(F3500) &  1.16 $\pm$ 0.07   &1.1$\pm$ 0.1 & 0.3 $\pm$ 0.4 &  1.12 $\pm$ 0.06 & 0.5 $\pm$ 0.3  \\
SFR$_{\mathrm{H\beta}}$ &  2.8 $\pm$ 0.6   &1.2$\pm$ 0.4 & 3.3 $\pm$ 0.5 & 2.4$\pm$ 0.7& 2.6 $\pm$ 0.4 \\
SFR$_{\mathrm{UV}}$ &  5.75 $\pm$ 0.06   &3.48$\pm$ 0.05 & 0.57 $\pm$ 0.05 & 6.54$\pm$ 0.05& 0.82 $\pm$ 0.05 \\
$\xi_{ion, NIRSpec}$ (erg$^{-1}$ Hz) & 25.0 $\pm$ 0.1 & 24.8 $\pm$ 0.2 & 26.1 $\pm$ 0.1 & 24.9 $\pm$ 0.1 & 25.8 $\pm$ 0.1\\
Burstiness (SFR$_{\mathrm{10 Myr}}$/SFR$_{\mathrm{100 Myr}}$) & 0.5 $\pm$ 0.2 & 0.4 $\pm$ 0.3 & 5.8 $\pm$ 0.2 & 0.4 $\pm$ 0.3 & 3.2 $\pm$ 0.2\\

\hline

\end{tabular}
\tablefoot{SFR$_{\mathrm{H\beta}}$ is derived assuming no extinction and the calibration from \citet{Theios19} for a metallicity of 20\% the solar value. Fluxes are in units of 10$^{-19} erg~s^{-1}~cm^{-2}$.}
    \label{tab:resolved_fluxes}
\end{table*}

\end{appendix}



\end{document}